\documentclass[10pt,a4paper,amsmath,amssymb,aps,prd,nofootinbib,notitlepage,twocolumn]{revtex4-2}
\usepackage{float}
\usepackage{lmodern}
\usepackage{booktabs}
\usepackage{graphicx}
\usepackage{dcolumn}
\usepackage{bm}
\usepackage{color}
\usepackage[normalem]{ulem}

\usepackage[caption=false]{subfig}

\usepackage[unicode=true,
 bookmarks=true,bookmarksnumbered=false,bookmarksopen=false,
 breaklinks=false,pdfborder={0 0 1},backref=false,colorlinks=true]
 {hyperref}

\definecolor{urlblue}{rgb}{0,0,0.9}\definecolor{linkblue}{rgb}{0,0,.8}\definecolor{linkgreen}{rgb}{0,0.45,0}\definecolor{linkpurple}{rgb}{0.7,0.0,0.4}\definecolor{linkorange}{rgb}{0.7,0.1,0.0}\AtBeginDocument{\hypersetup{
linkcolor=linkblue,
citecolor=linkorange,
urlcolor=urlblue}}\definecolor{urlblue}{rgb}{0,0,0.9}\definecolor{linkblue}{rgb}{0,0,.8}\definecolor{linkgreen}{rgb}{0,0.45,0}\definecolor{linkpurple}{rgb}{0.7,0.0,0.4}\definecolor{linkorange}{rgb}{0.7,0.1,0.0}
\AtBeginDocument{\hypersetup{
linkcolor=linkblue,
citecolor=linkorange,
urlcolor=urlblue}}

\definecolor{commentblue}{rgb}{0,0,.85}

\usepackage[mathlines]{lineno}

\newcommand{\dd}{\textrm{d}}

\newcommand{\tC}{\widetilde{C}}
\newcommand{\tN}{\widetilde{N}}

\newcommand{\betad}{\boldsymbol{\beta}^{\rm D}}

\newcommand{\nhat}{\bm{\hat{n}}}
\newcommand{\nhatp}{\bm{\hat{n}^{\prime}}}
\newcommand{\dint}{\boldsymbol{\Delta}_{1,{\rm int}}}
\newcommand{\done}{\boldsymbol{\Delta}_{1}}

\graphicspath{{prd-plots/}}

\begin{document}

\title{Disentangling Doppler modulation, aberration \\ and the temperature dipole in the CMB}

\author{Pedro da Silveira Ferreira}
\email{pferreira@astro.ufrj.br}
\thanks{Both authors contributed equally}
\affiliation{Observat\'orio do Valongo, Universidade Federal do Rio de Janeiro, 20080-090, Rio de Janeiro, RJ, Brazil}
\author{Miguel Quartin}
\thanks{Both authors contributed equally}
\affiliation{Observat\'orio do Valongo, Universidade Federal do Rio de Janeiro, 20080-090, Rio de Janeiro, RJ, Brazil}
\affiliation{Instituto de F\'isica, Universidade Federal do Rio de Janeiro, 21941-972, Rio de Janeiro, RJ, Brazil}

\date{\today}

\begin{abstract}
    An observer in relative motion to the Cosmic Microwave Background (CMB) rest frame is sensitive to both aberration and Doppler effects. Both effects introduce similar but non-identical off-diagonal couplings in the spherical harmonic coefficients. The CMB temperature dipole may have additional contributions from an intrinsic component, which in turn produces different aberration and Doppler couplings. Moreover, the standard conversion from intensity measurements into temperature also introduces spurious Doppler-like couplings. In order to learn about the intrinsic dipole it is therefore important to measure both aberration and Doppler couplings in an independent manner while also removing the spurious contributions from unit conversion, which are degenerate with the dipole. Here we present a pipeline to measure the Doppler and aberration signal independently from each other and from the dipole itself. We also consider realistic beaming, noise and mask effects. Our pipeline results in independent and unbiased estimators which have uncertainties only $\simeq 20\%$ larger than the simple theoretical expectations.  We discuss the achievable precision in each measurement for Planck 2018, and also forecast them for future ground-based experiments with the Simons Observatory and CMB-S4. An alternative pipeline is presented in order to cross-check results and improve robustness.
\end{abstract}

\maketitle


\section{Introduction}\label{sec:intro}

The first clear measurement of the Cosmic Microwave Background (CMB) temperature dipole dates back almost 50 years~\cite{Henry1971}. It has two main components: the orbital dipole, due to the motion around the Sun, and the solar dipole, due to the motion of the solar system with respect to the CMB rest frame. The former has a predictable yearly modulation and therefore was used to calibrate both WMAP and Planck maps~\cite{Jarosik:2010iu,Akrami:2018vks}. The latter, measured after removal of the orbital contribution, is the largest CMB anisotropy, with an amplitude of $3.36208 \pm 0.00099$ mK~\citep{Akrami:2018vks}. This is $\sim100$ larger than the anisotropies in the other multipoles $\ell$ until $\ell\sim 1000$ (in higher $\ell$s, the Silk damping effect makes fluctuations increasingly smaller).

The high amplitude of the temperature dipole is the main reason for it to be often fully credited to the proper motion between the solar system and the CMB rest frame. If one assumes that the entire dipole has such an origin, the inferred relative velocity is $(369.82 \pm 0.11)$~km/s. This velocity estimate is of importance beyond early universe physics and the CMB, and it is regularly used in astronomy in order to convert observed redshifts into heliocentric redshifts. There is however no reason to assume that there cannot be an intrinsic component to the dipole. Concrete alternatives to the kinematic scenario were discussed as already 3 decades ago by~\cite{Paczynski1990}, which showed that a large local void could also explain the dipole. This particular scenario was further investigated by e.g.~\cite{Tomita:1999rw} and~\cite{Quartin:2009xr}. A ``tilted universe scenario'' composed of a superhorizon isocurvature perturbation was proposed in~\cite{Turner_1992}. An inflationary model which produces similar results was proposed by~\cite{Langlois:1996ms}.

The separation of primordial and kinematic effects in the CMB is not straightforward as in the case of adiabatic perturbations there is a degeneracy between the Doppler effect and the primordial perturbations. As discussed in detail in~\cite{Roldan:2016ayx}, however, this degeneracy can be broken by measuring the Doppler-like and aberration-like couplings in the CMB. In the case of a peculiar velocity these couplings must be present and with well-determined coefficients, as discussed in \cite{Amendola:2010ty} and~\cite{Notari:2011sb}~\cite[see also][for a review]{Yasini:2019ajn}. In other scenarios, one of both of these couplings can differ. Therefore if one measures the dipole, Doppler and aberration effects independently one can learn about the intrinsic CMB dipole and test the hypothesis that the CMB dipole is mostly due to a peculiar velocity. It also allows for a more model-independent measurement of our peculiar velocity, which has implications in inferring the cosmological redshift of all sources.

The aberration and Doppler couplings of the CMB were shown to be detectable by Planck by~\cite{Kosowsky:2010jm} and~\cite{Amendola:2010ty}, and subsequently measured by~\cite{Aghanim:2013suk} following an estimator proposed in~\cite{Hanson:2009gu}. These couplings were also shown to affect the measurements of some CMB anomalies such as the dipolar modulation (and the connected hemispherical asymmetry)~\cite{Quartin:2014yaa} and the quadrupole anomalies~\cite{Notari:2015kla}. On the other hand, if the primordial map is Gaussian it was shown not to affect $f_{\rm NL}$ measurements~\cite{Catena:2013qd}. Alternative methods to measure these couplings were also recently proposed in~\cite{Aurich:2021wwx}. In~\cite{Notari:2015daa} it was shown  that similar couplings arise, with higher significance than the temperature couplings, as a leakage of temperature correlations into thermal Sunyaev-Zeldovich (tSZ) maps, which can be measured by cross-correlating tSZ and temperature ($T$) maps. This cross-correlation was recently detected with significance above $5\sigma$ by~\cite{Akrami:2020nrk}, but this effect is physically \emph{completely} degenerate with the basic dipole measurement and serves only as a cross-check of the model and data. Finally, exploiting the frequency-dependence of these couplings in the low-multipoles of the CMB was proposed as a way to separate the intrinsic and kinematic components of the dipole~\cite{Yasini:2016dnd}.

The Planck measurements of aberration and Doppler in~\cite{Aghanim:2013suk} made use of only the 143GHz and 217GHz channels in Planck 2013 data and measured $v = 384 \,\rm{km/s} \pm 78 \, \rm{km/s \, (statistical)}\pm 115$ km/s (systematic). As explained in that work this measurement however did not distinguish from possible intrinsic contributions and is at least partly degenerate with the standard dipole measurements since the spurious Doppler couplings from the conversion of intensity into temperature was not removed. Similar measurements have been very recently performed in Planck 2018 data, yielding a result of $v=(298.5 \pm 65.6)$~km/s in a direction compatible with the dipole~\cite{Saha:2021bay}. In this work we propose several improvements to the methodology of~\cite{Aghanim:2013suk}. First of all, we propose to remove the couplings which bring no new information with respect to the dipole. Second, we propose to remove biases in the estimators of aberration and Doppler by simulating the effects using many combinations of orientations in the simulations.
This ensures that no a priori information on the direction of either effect is assumed. Third, for Planck we propose using the final component separated maps of SMICA and NILC instead on single frequency maps. Finally, we discuss also the benefits of adding $E$-mode polarization maps ($E$) to both aberration and Doppler measurements. We discuss the achievable precision in both measurements for Planck 2018 using this pipeline and also forecast the precision for future ground-based experiments with the Simons Observatory~\cite{Ade:2018sbj} and CMB-S4~\cite{Abazajian:2016yjj}.

These proposed improvements result not only in much smaller systematic errors but also in better precision. More importantly, they hinge on less assumptions. They were also applied to real Planck 2018 data in our companion paper~\cite{Ferreira:2020aqa}, where for the first time we are able to put an upper bound on the intrinsic CMB dipole and find that its amplitude must be $< 3.7$ mK at 95\% confidence level. An estimate of our peculiar velocity with the CMB without assuming a negligible intrinsic component was also made, resulting in
$v = (300\pm 99)$ km/s with $(l,b) =(276\pm 32, \,51\pm 19)^\circ$ [SMICA], and  $v~=~(300~\pm~100)$ km/s with $(l,b) =(280\pm 32, \,50\pm 20)^\circ$ [NILC].

\section{Independent Estimators for Doppler and Aberration}\label{sec:estimator}

As discussed in \cite{Challinor:2002zh} and \cite{Amendola:2010ty}, the primordial temperature $T(\hat{\bm{n}})$ on the direction $\hat{n}$ seen by an observer with a peculiar velocity $\bm{\beta}$ turns into the boosted $T^{\prime}(\hat{\bm{n}}^{\prime})$ following the equation
\begin{align}\label{eq:doppler}
    T^{\prime}(\hat{\bm{n}}^{\prime}) & =\frac{T\left(\hat{\bm{n}}\right)}{\gamma^{\rm D}\left(1-\bm{\beta}^{\rm D} \cdot \hat{\bm{n}}\right)} \;,
\end{align}
with the aberrated direction $\nhatp$ given by
\begin{equation}
    \nhatp  =  \frac{\bm{\hat{n}}\cdot\bm{\hat{\beta}}^{\rm A}+\beta^{\rm A}}{1+\bm{\hat{n}}\cdot \bm{\beta}^{\rm A}}\,\bm{\hat{\beta}}^{\rm A}+\frac{\left[\bm{\hat{n}}-(\bm{\hat{n}}\cdot\bm{\hat{\beta}}^{\rm A})\bm{\hat{\beta}}^{\rm A}\right]}{\gamma^{\rm A}(1+\bm{\beta}^{\rm A}\cdot\bm{\hat{n}})}\,,
    \label{eq:aberration_n}
\end{equation}
where the index A stands for aberration and D for Doppler, and $\gamma^{\rm X}= [1-(\beta^{\rm X})^2]^{-1/2}$. In the traditional case where our peculiar velocity is the only source of aberration-like and Doppler-like signals one has $\bm{\beta}^{\rm A}=\bm{\beta}^{\rm D} = \bm{\beta}$. In this work instead we separate explicitly the sources of Doppler and aberration in order to further test the peculiar velocity assumption. We will therefore allow also for the case in which $\bm{\beta}^{\rm A}\neq \bm{\beta}^{\rm D}$. This allows in principle the measurement of a non-adiabatic intrinsic dipole in the CMB, as discussed in detail in~\cite{Roldan:2016ayx}. Doppler and aberration due to a boost likewise affect the polarization maps and as shown in~\cite{Notari:2011sb} for Planck there should be non-negligible information in polarization, both for the two point estimators using only $E$ maps ($EE$) and cross-correlating with temperature ($TE$+$ET$).

The above can then be expanded on spherical harmonics. For a given CMB map the coefficients $a_{\ell m}$ can be expanded in a first order approximation in $\beta$. For $\beta \sim 10^{-3}$ such an expansion works surprisingly well all the way up to $\ell \sim 3000$~\citep{Chluba:2011zh,Notari:2011sb}. We therefore proceed this way, which means that Doppler and aberration couple the $\ell$, $\ell+1$ components but not other $\ell$, $\ell+n$. We also separate the aberration and Doppler contributions in the final $a_{\ell m}$s:
\begin{equation}
    a^{\prime}_{\ell m} = a^{\rm{Prim}}_{\ell m} + a^{\text{A}}_{\ell m} + a^{\text{D}}_{\ell m} \,,
    \label{eq:alm_exp}
\end{equation}
where $a^{\rm{Prim}}_{\ell m}$ are the primordial coefficients. It is convenient to define the quantity
\begin{align}\label{eq:Glm}
    {}_sG_{\ell\, m} \;\equiv\; \sqrt{\frac{\ell^2-m^2}{4\ell^2-1}\left[1-\frac{s^2}{\ell^2}\right]},
\end{align}
where $s=0$ for temperature and $s=2$ for polarization. We can then write, first considering $\bm{\beta}$ only in the $z$ direction
\begin{align}
    a_{\ell m}^{\rm A} & = c_{\ell m}^{\rm A,-}a_{\ell-1\, m}^{\rm Prim}+c_{\ell+1 m}^{\text{A},+}a_{\ell+1\, m}^{\rm Prim},  \label{eq:almab}\\
    a_{\ell m}^{\rm D} & = c_{\ell m}^{\rm D}a_{\ell-1\, m}^{\rm Prim}+c_{\ell+1 m}^{\rm D}a_{\ell+1\, m}^{\rm Prim} ,
    \label{eq:almdopp}
\end{align}
with
\begin{equation}
    c_{\ell m}^{\rm D} = -\beta^{\rm D}_{z}{}_sG_{\ell\, m} \quad \mathrm{and}\quad c_{\ell m}^{\rm A,\pm} = \beta^{\rm A}_{z}(1\pm\ell) {}_sG_{\ell\, m}\,.\nonumber
    \label{ccabdopp}
\end{equation}

For the case of only a peculiar velocity ($\bm{\beta}^{\rm A} = \bm{\beta}^{\rm D} \equiv \bm{\beta}$), following~\cite{Amendola:2010ty} one defines the quantity
\begin{equation}
    f_{\ell m} \equiv \operatorname{\mathbb{R}e}\big[a^*_{\ell\, m}a_{\ell+1\, m}\big] \,,
\end{equation}
the theoretical expectation of which is
\begin{align}
     \langle f_{\ell m}\rangle = \big[c_{\ell+1 m}^{\rm A,-} + c_{\ell+1 m}^{\rm D}\big] \!C_{\ell} +\! \big[c_{\ell+1 m}^{\rm A,+} +  c_{\ell+1 m}^{\rm D}\big] \!C_{\ell+1}.
    \label{fth2}
\end{align}
It is then convenient to define $\hat{f}_{\ell m}^{\rm TH} \equiv \langle f_{\ell m}\rangle / \beta_z $ which is independent of $\beta_z$ at first order and use the following estimator
\begin{equation}
    \hat{\beta}_{z}=\left(\sum_{\ell,m}\frac{f_{\ell m}^{\rm obs}\hat{f}_{\ell m}^{\rm TH}}{{\mathfrak C}_{\ell}{\mathfrak C}_{\ell+1}}\right)\left(\sum_{\ell,m}\frac{(\hat{f}_{\ell m}^{\rm TH})^{2}}{{\mathfrak C}_{\ell}{\mathfrak C}_{\ell+1}}\right)^{-1},
    \label{eq:estimatorz}
\end{equation}
where ${\mathfrak C}_{\ell} \equiv (C_{\ell}+N_{\ell})$  is the sum of the signal and noise spectra.

In the general case where $\bm{\beta}$ has all cartesian components, the above is generalized following Appendix B of~\cite{Amendola:2010ty}. In particular it involves the two point functions $a_{\ell m}a_{\ell+1 m\pm1}$.   One can also make use of the Wigner rotation matrix to rotate (using the $a_{\ell m}$s) a vector in the $z$-axis to any other point in the sphere and \emph{vice-versa}.

The above estimator does not consider the important effects of a sky mask. Moreover, it does not account for anisotropic noise and other systematic effects. Applying the mask $W$ and adding noise we get
\begin{equation}
    \tilde{a}_{\ell m} = \sum_{\ell'm'} a_{\ell'm'} K_{\ell m\ell'm'}[W] + \sum_{\ell''m''} a^{\rm noise}_{\ell''m''} K_{\ell m\ell''m''}[W] \; ,
    \label{eq:almmaskNew}
\end{equation}
where  $K_{\ell ml'm'}[W]$ is the mask kernel and $a^{\rm noise}_{\ell m}$ the components of the noise map. Considering $\ell_{\rm max} \sim 2000$ it is computationally unfeasible to obtain the inverse matrix of $K_{\ell ml''m''}[W]$ to remove the mask effect. Therefore we instead consider as a first step the effect of the mask only on the angular power spectra in what we call our \emph{baseline} estimator. For the noise, we also consider initially only its angular power spectrum, ignoring its anisotropies. We then take into account the anisotropies of both mask and noise by performing realistic simulations in HEALPix in which the full mask is added and noise included using the set of Planck dx12 noise simulations for the corresponding map-making procedure. With these simulations we apply additive and multiplicative corrections to the baseline estimator. This is described in more detail in Section~\ref{sec:bias}.

The isotropic effect of the mask is a change on the angular power spectrum $C_\ell$:
\begin{equation}
    \tC_{\ell_1} = \sum_{\ell_2} M_{\ell_1\ell_2} C_{\ell_2} \quad {\rm and}\quad \tN_{\ell_1} = \sum_{\ell_2} M_{\ell_1\ell_2} N_{\ell_2}\,,
    \label{eq:pseudo_cl_master}
\end{equation}
where a tilde is used when including the mask.  $M_{\ell_1\ell_2}$ is called the MASTER correlation matrix and is discussed in detail in~\citep{Hivon:2001jp,Pereira:2010dn}. The masked angular spectra including noise are then represented by ${\tilde{\mathfrak C}}_{\ell} \equiv \tC_{\ell} + \tN_\ell$, and this is the quantity we employ on Eq.~\eqref{eq:estimatorz} for our baseline estimator discussed above.

As discussed previously, here we are interested also in the case $\bm{\beta}^{\rm A}\ne\bm{\beta}^{\rm D}$. This encourages the definition of
\begin{equation}
    \beta_z^{\rm X} \hat{f}_{\ell m}^{\rm TH,X} \,\equiv\, c_{\ell+1 m}^{\rm X,-} C_{\ell} + c_{\ell+1 m}^{\rm X,+} \!C_{\ell+1} ,\label{fth}
\end{equation}
for X = A or D and ``D,+'' = ``D,$-$'' = D. However, simply rewriting the estimator above for $\bm{\beta}^{\rm A}$ and $\bm{\beta}^{\rm D}$ separately would not consider the correlations between the aberration and Doppler signals. In fact, since both effects introduce $\ell, \ell+1$ correlations, they cannot be disentangled exactly.
The most precise way to solve this problem would be to minimize a $\chi^2$ numerically for independent $\bm{\beta}^{\rm D}$ and $\bm{\beta}^{\rm A}$ for each vector component:
\begin{equation}
    \chi^{2}(\beta^{\mathrm{A}}_i,\beta^{\mathrm{D}}_i)=2\sum_{\ell,m}\frac{\left[f_{\ell m}^{\rm obs}-\beta^{\mathrm{A}}_i\hat{f}_{\ell m}^{\rm TH,\mathrm{A}}-\beta^{\mathrm{D}}_i\hat{f}_{\ell m}^{\rm TH,\mathrm{D}}\right]^{2}}{{\mathfrak C}_{\ell}{\mathfrak C}_{\ell+1}}\;.\label{chi2betaAD}
\end{equation}
That is nevertheless computationally very demanding considering the number of simulations we make use of to correct the bias of masking and anisotropic noise (see Section~\ref{sec:bias}). The Planck Collaboration separated both aberration and Doppler signals using orthogonalized weight matrices~\cite{Aghanim:2013suk}. A simpler solution is to just use independent estimators for A and D following Eq.~\eqref{eq:estimatorz} and to remove the correlation \emph{a posteriori} on the data analysis. We adopt this computationally simple, but very effective approach, considering the correlation as a linear effect which is described in detail on Section~\ref{ap:correlation} and which allows an approximately independent measurement of both aberration and Doppler. We find that this approach is appropriate even for higher precision levels than the ones achievable on present data.

We also consider the more traditional case in which one assumes \emph{a priori} that $\bm{\beta}^{\rm A} \equiv \bm{\beta}^{\rm D} \equiv \bm{\beta}^{\rm B}$ as in a standard boost~(B) transformation. This corresponds to a boost due to a peculiar velocity in the standard slow-roll inflation scenario, which does not allow for any additional source of Doppler-like and aberration-like couplings. This in general leads to a higher significant detection of these couplings if indeed there are no other sources of such couplings besides a regular boost. On the other hand such a measurement reduces to a simple cross-check, as assuming no other sources of such couplings means that all physical information is already encoded in the high-precision observation of the temperature dipole. The estimated uncertainties in $\beta^{\rm A}$, $\beta^{\rm D}$ and $\beta^{\rm B}$ of both ideal (without mask and noise) and realistic simulated cases are shown in Section~\ref{app:ap_ideal_stat}.

\section{The Dipole Distortion effect on the estimators}\label{sec:dipoledoppler}

The Planck collaboration does not provide the exact temperature map from their measurements. Instead it converts intensity maps on different frequencies $\nu$ into a temperature map using a linearized relation between both quantities. Defining the quantity $\delta T / T_0$ as the first-order dimensionless temperature anisotropies for $\ell \geq 2$, this procedure introduces distortions in the CMB maps of the order $\sim \beta \delta T / T_0$, which are not due to any primordial process. This was originally pointed out by~\cite{Kamionkowski:2002nd} in the particular case of the quadrupole and developed in the general case by~\cite{Notari:2015daa} \citep[see also][]{Mukherjee:2013zbi}. We summarize the main points below.

The bolometric specific intensity $I(\nu,\boldsymbol{\hat{n}})$ on the CMB rest frame is given by:
\begin{equation}
   I(\nu,\boldsymbol{\hat{n}}) = \frac{h}{c^2}\frac{2 \nu^3}{e^{\frac{h \nu }{k_B T(\boldsymbol{\hat{n}})}}-1} \,. \label{eq:Int}
\end{equation}
We Taylor expand around $T_0$ to first order, decomposing $T(\boldsymbol{\hat{n}})=T_0+\Delta T(\boldsymbol{\hat{n}})$, and get
\begin{equation}
    \delta I(\nu,\boldsymbol{\hat{n}})\,\approx\, \frac{h}{c^2}\frac{2 \nu ^4   e^{\frac{\nu }{\nu_0}}}{T_0^2 \left(e^{\frac{\nu }{\nu_0}}-1\right)^2}   \, \delta T(\boldsymbol{\hat{n}})
    \,\equiv\, K(\nu) \, \frac{\Delta T(\boldsymbol{\hat{n}})}{T_0} \,,
    \label{eq:lin-temp-1st-order}
\end{equation}
where $\nu_0 \equiv k_{B}T_0 / h = (56.79\pm 0.01)$ GHz~\cite{Fixsen:2009ug}.

The default approach since WMAP is to \emph{define} temperature fluctuations $\Delta T$ as a linear correction to the intensity fluctuations $\delta I$: $\Delta T/ T_0(\bm{\hat{n}})\big|_{\rm linear} \equiv \delta I(\nu,\bm{\hat{n}})/K(\nu)$, which following~\cite{Notari:2015kla} we will refer to as the \emph{linearized temperature} $L(\bm{\hat{n}})\equiv \delta I(\nu,\bm{\hat{n}})/K(\nu)$. If we now extend the expansion to second order we get\footnote{This corrects a typo in~\cite{Notari:2015daa} where the $-1$ was missing.}
\begin{align}\label{eq:lin-temp-2nd-order}
    L(\nu,\bm{\hat{n}}) \,=\, \frac{\Delta T(\bm{\hat{n}})}{T_0} + \left( \frac{\Delta T(\bm{\hat{n}})}{T_0} \right)^2 \big[Q(\nu)-1\big]\,,
\end{align}
where
\begin{equation}
	Q(\nu) \,\equiv\, \frac{\nu}{2\nu_0}  \coth\left[\frac{\nu}{2\nu_0}\right] .
\end{equation}
The second order term shows that higher-order black-body distortions would appear.

Following~\citet{Notari:2015daa} we expand the linearized temperature $L$ into terms which contain perturbations ${\cal O}\big(10^{-8}\big)$ or higher. These include terms ${\cal O}\big(\beta^2\big)$ and ${\cal O}\big( \beta \delta T/T_0 \big)$, but not ${\cal O}\big(\beta^3\big)$ or ${\cal O}\big([\delta T/T_0]^2\big)$. As discussed in~\citep{Roldan:2016ayx} and in our companion Letter~\cite{Ferreira:2020aqa}, we can gain information on the nature of the intrinsic CMB dipole by confronting the aberration, Doppler and dipole measurements. In particular, the Doppler couplings contain contributions from this intrinsic component, but the amplitude of these couplings are in general not the same as the temperature dipole. So it is convenient to write for the observed dipole $\done\equiv \bm{\beta}+\dint$, where $\bm{\beta}$ is our velocity and $\dint$ is the intrinsic dipole; and for the Doppler couplings coefficient $\betad = \bm{\beta}+\bm{\alpha}$, where $\bm{\alpha} \equiv (1+\alpha^{\rm NG}) \dint$~\cite{Ferreira:2020aqa} are the intrinsic Doppler couplings,\footnote{Note that we are now adopting is a slightly different notation than in Eq.~\eqref{eq:doppler}, but which is more in line with what was demonstrated in~\cite{Roldan:2016ayx}, i.e. that $\bm{\alpha}$ affects only the Doppler couplings and not the full Lorentz boost.}  which depend on the primordial anisotropy scenario. The complete temperature anisotropy $\Delta T$ is therefore written as $\Delta T(\nhat) = \done T_0 + \delta T(\nhat) + (\bm{\alpha}\cdot\nhat)\delta T(\nhat)+ {\cal O}\big(10^{-9}\big)$. Including the effect of aberration and Doppler on $I(\nu,\bm{\hat{n}})$, one gets:
\begin{equation}
   I^\prime(\nu^{\prime},\nhatp) = \frac{h}{c^2}\frac{2 {\nu^{\prime}}^3}{e^{\frac{h \nu }{k_B T(\nhat)}}-1} \,, \label{eq:Int_boosted}
\end{equation}
where $\nu = \gamma (1-\bm{\beta} \cdot \bm{\hat{n}}) \nu^{\prime}$. We thus arrive at
\begin{equation}
   I^\prime(\nu^{\prime},\nhatp) = \frac{2  h}{c^2}
   \frac{{\nu^{\prime}}^3}{e^{\frac{\nu^{\prime} \gamma (1-\bm{\beta} \cdot \nhat )}{\nu_0 \left(1+\frac{\delta T(\nhat)}{T_0} + \dint \cdot \nhat + \frac{\delta T(\nhat)}{T_0} \bm{\alpha} \cdot \nhat   \right)}
   }-1} \, .\label{eq:I_boost_full}
\end{equation}

Expanding to the order discussed above and dividing by $K(\nu^\prime)$ we arrive at:
\begin{align}\label{eq:L-2nd-order}
    L(\nu^{\prime},\nhatp)\! \,=\,  &\!-\frac{1}{2}\beta^2+\frac{1}{3}\Delta_1^{2}Q(\nu^{\prime}) + \done\cdot\bm{\hat{n}} \nonumber \\ &\!+\Delta_1^{2}Q(\nu^{\prime})\left[(\bm{\hat{\Delta}_1}\cdot\bm{\hat{n}})^2 - \frac{1}{3}\right]       \nonumber \\
     \;&\!+(\bm{\beta}\cdot\bm{\hat{n}}) (\done\cdot\bm{\hat{n}})- (\done\cdot\bm{\hat{n}})^2   \nonumber \\
    & \!+  \frac{\delta T(\bm{\hat{n}})}{T_0}
    \!\Big[\betad\!\cdot\!\bm{\hat{n}}  +  2\done\!\cdot\!\bm{\hat{n}} \big(Q(\nu^{\prime})\!-\!1\big) \Big] \nonumber \\
    & \!+ \frac{\delta T(\bm{\hat{n}})}{T_0}
    +  \beta^{\rm A}  \frac{\delta T_{ab}(\bm{\hat{n}})}{T_0}
    + {\cal O}\big(10^{-9}\big)\,.
\end{align}
This is the most general way to decompose the terms such as to isolate all effects that carry no new information apart from the dipole, and represents an extension over the results in~\cite{Notari:2015daa}.
The first two terms are contributions to the monopole, and in particular the second is sometimes called the $y$-type monopole~\citep{Chluba:2004cn}. The third term is the standard CMB dipole. The terms on the second and third lines are the Doppler-quadrupole terms. They carry a frequency dependence, originally discussed by~\citet{Kamionkowski:2002nd}. Here we expanded the Doppler quadrupole into 3 terms because we allow for an intrinsic dipole; when it is assumed to be zero, only the first term remains. The next terms are the ones we are interested in this paper: they are the Doppler modulation due to $\betad$ (a combination of our peculiar velocity and intrinsic contributions) and the Dipole Distortions (DD), which we discuss below. Finally the aberration effect in the anisotropies
are described by $\beta^{\rm A} \delta T_{ab}/T_0$, which includes our velocity and a possible contribution to aberration from an intrinsic dipole~\citep{Roldan:2016ayx,Ferreira:2020aqa}, but note that the total effect is expected to have an amplitude similar to $\beta$. This is also what we find in the real data in the companion Letter~\cite{Ferreira:2020aqa}.

In what follows for simplicity of notation we will omit the primes from $\nu$ and $\bm{\hat{n}}$ since we will always be referring to quantities in the rest frame of the observer.

As discussed in detail by~\cite{Notari:2015daa}, the crucial point is that the DD do not add any new information not already contained in the very precise measurements of $\done$. Nevertheless, they leave an imprint in the data which is degenerate with the Doppler modulations. Therefore, in order to make measurements of Doppler and aberration which are independent from the basic dipole measurements one needs to remove this contribution in the analysis. This was first realized by the Planck Collaboration in their measurement of the Doppler modulations~\citep{Aghanim:2013suk}, but they did not attempt to make this separation. Their Doppler modulation measurements are thus driven in part by the dipole itself.

We so far discussed the DD only in terms of the temperature maps, but the same effect is also present in the $E$ and $B$ polarization maps~\citep{Mukherjee:2013zbi}. In the $B$ maps much higher precision is needed to get good measurements of this effect than what is expected in the near future, so we will ignore it. But this is not the case for the $E$ maps. We therefore also need to remove the DD from the $E$ map.

In order to remove the redundant DD contribution from our Doppler measurements we need to estimate the effective weight of the frequency dependent contributions arising from $Q(\nu)$ on a given mapmaking procedure. Here we will therefore only consider the Planck mapmaking methods that are mostly based on linear combinations of single-frequency maps on harmonic space which make it possible to calculate the effective DD($\ell$) contribution in a straightforward manner. In particular, from the four main CMB maps we will restrict ourselves to only two: SMICA and NILC. These pipelines rely on the sum of single-frequency maps using weights that varies with $\ell$ in a way that optimizes the CMB signal extraction.

\begin{figure}
    \centering
    \includegraphics[width=0.9\columnwidth]{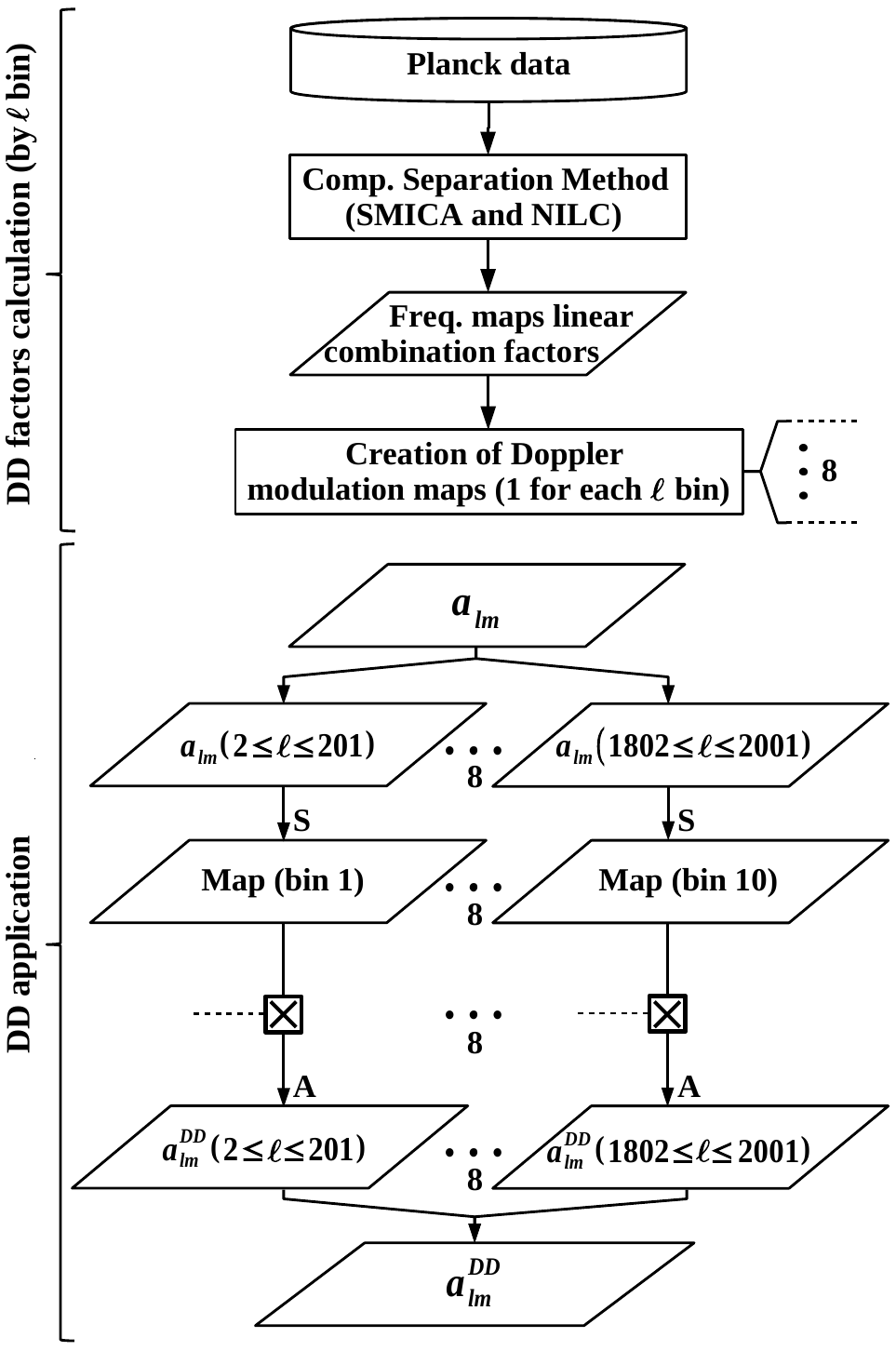}
    \caption{Doppler Distortion pipeline. For each temperature and polarization map an equivalent Doppler modulation map is created for each multipole bin with the effective Doppler modulation by applying the DD directly on pixel space [see text].     We use 10 bins of $\Delta\ell=200$.  \textbf{A} and \textbf{S} indicate the use of HEALPix routines \textit{anafast} and \textit{synfast}, respectively.
    \label{fig:dd_pipeline}}
\end{figure}

We can now define the DD factor ${\rm DD}^\mathrm{M}_\ell$ for each mapmaking method as
\begin{equation}
    {\rm DD}^{\rm M}_\ell \equiv  \sum_{\nu} 2 X^{\mathrm{M}}_{\ell,\nu} \big[Q(\nu)-1\big] \,,
    \label{eq:dd_by_l}
\end{equation}
where $X^{\mathrm{M}}_{\ell,\nu}$ is the weight of each multipole in the frequency channel $\nu$ of map M. For each $\ell$, the sum of all single frequency components have to obey
\begin{equation}
	\sum_{\nu}{X^{\mathrm{M}}_{\ell,\nu}} = 1 .
\end{equation}
The SMICA's $X^{\rm SMICA}_{\ell\nu}$ and NILC's $X^{\mathrm{NILC}}_{\ell\nu}$ weights are defined on \cite{Akrami:2018mcd} for temperature as\footnote{Considering debeaming, calibration and the mask, as per the SMICA propagation code available at \url{https://wiki.cosmos.esa.int/planck-legacy-archive/index.php/SMICA_propagation_code}\label{refnote}}
\begin{align}
    X^{\mathrm{ SMICA},T}_{\ell\nu} & = \frac{ W^{\mathrm{Full ,}T}_{\ell\nu}}{b_{\ell\nu}c_{\nu}}  +
    \mathcal{P(\ell)}
    \left[ \frac{W^{\mathrm{High,}T}_{\ell\nu}}{b_{\ell\nu}}
    -   \frac{ W^{\mathrm{Full,}T}_{\ell\nu}}{b_{\ell\nu}c_{\nu}}  \right], \\
    {X^{\mathrm{NILC,}T}_{\ell\nu}} & = \frac{\sum\limits_{\rm band}{ W^{\mathrm{NMW},T,\mathrm{band}}_{\nu} {h^{\rm band}_\ell}}}{\sum\limits_{\rm band}{h^{\rm band}_\ell}}\,.
\end{align}
In the above $\mathcal{P(\ell)}$ is a linear operator that apply a high-pass filter with a multiplicative factor,\footnote{This factor is the sky fraction ($f_{\rm sky}$) of the transition mask used to combine $W^{\mathrm{Full}}$ and $W^{\mathrm{High}}$ considering different weights for regions near and far from the galactic plane on SMICA 2018 pipeline. This is useful to compute the effective weight for a full map and were used to compare our pipeline with Planck's dx12 simulations. When using masked maps $X^{\rm SMICA}_{\ell\nu} \simeq W^{\mathrm{High}}_{\ell\nu}/b_{\ell\nu}$ for $\ell\geq150$ (equal if the mask covers all the transition region).}
$W^{\mathrm{Full}}_{\ell\nu}$ and $W^{\mathrm{High}}_{\ell\nu}$ are the weights for each frequency considering all frequencies (Full) or only high frequencies (100, 143, 217, 353, 545 and 857 GHz), $b_{\ell\nu}$ is the beam function and $c_{\nu}$ is a calibration factor.
$W^{\mathrm{NMW,band}}$ are the mean weights for each single-frequency component for each band, where the needlets ${h^{\rm band}_\ell}$ define the extension and weight of the bands. For polarization $E$ channel, only 30, 44, 70, 100, 143, 217, 353 GHz frequencies are used and SMICA weights are defined on \cite{Akrami:2018mcd} as
\begin{align}
    X^{\mathrm{ SMICA}, E}_{\ell\nu} & = \frac{ W^{\mathrm{Full} , E}_{\ell\nu}}{b_{\ell\nu}c_{\nu}}  , \\
    {X^{\mathrm{NILC} , E}_{\ell\nu}} & = \frac{\sum\limits_{\rm band}{ W^{\mathrm{NMW},E, \mathrm{band}}_{\nu} {h^{\rm band}_\ell}}}{\sum\limits_{\rm band}{h^{\rm band}_\ell}}\,.
\end{align}
The DD generates an effective Doppler-like modulation with an effective velocity     $\bm{\beta}^{\rm DD,M}(\ell) \,\equiv\, {\rm DD}^{\rm M}_{\ell}\bm{\Delta}_{1}\,,$ this modulation is of the order of $\sim 300-600 \,{\rm km/s}$, pointing on the dipole direction. Since it is degenerate with any Doppler-like signal, results will be biased towards the dipole direction if this is not removed. We therefore built a pipeline which allows an accurate removal of the DD. Since this is an important issue, we built an alternative pipeline to cross-check the DD removal. Both pipelines are discussed in Section~\ref{sec:simulations}. The complete process of reproducing the DD is summarized in Figure~\ref{fig:dd_pipeline}.

\begin{figure}
    \includegraphics[trim={0cm 2.3cm 0cm .4cm}, clip, width=.95\columnwidth]{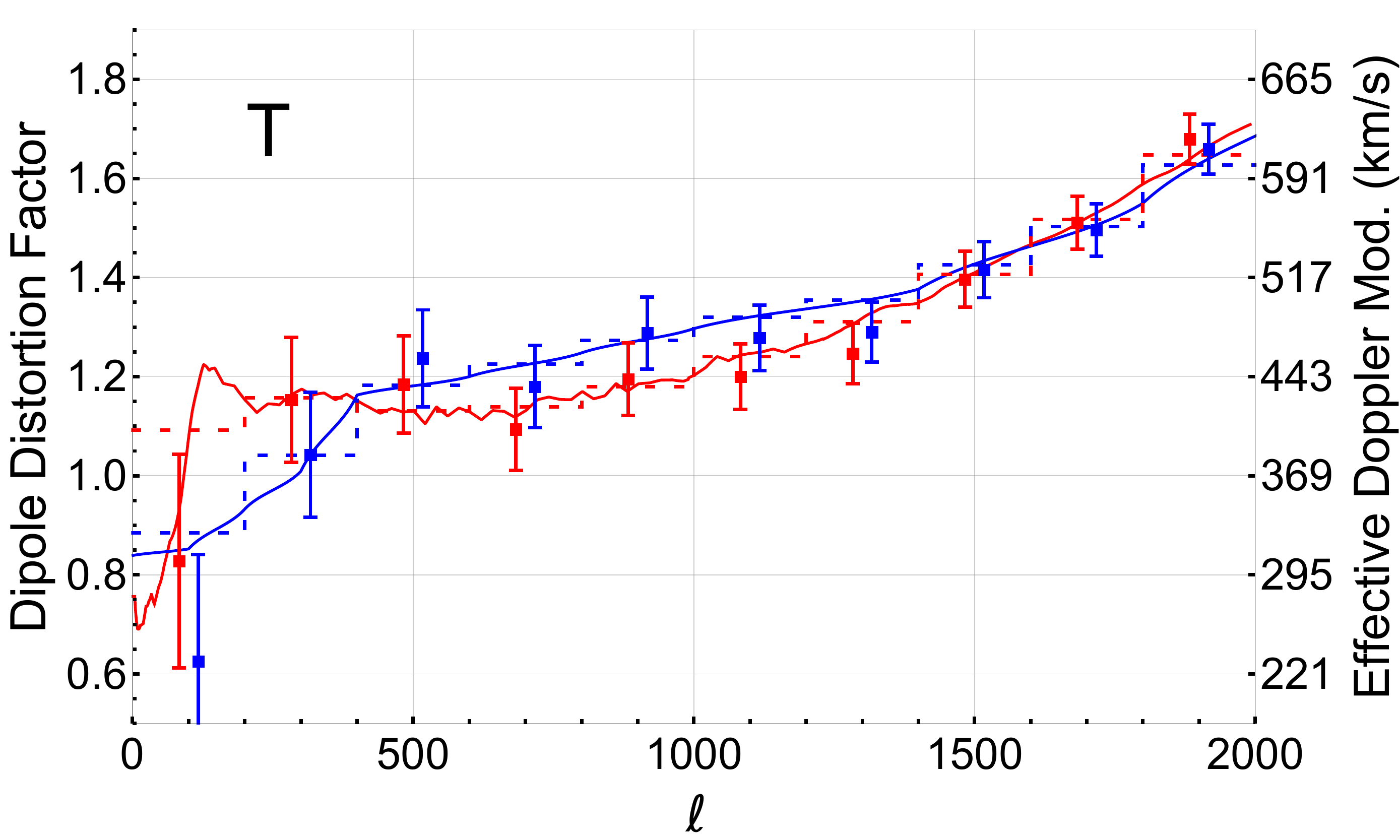}
    \includegraphics[trim={0cm 0cm 0cm .2cm}, clip, width=.95\columnwidth]{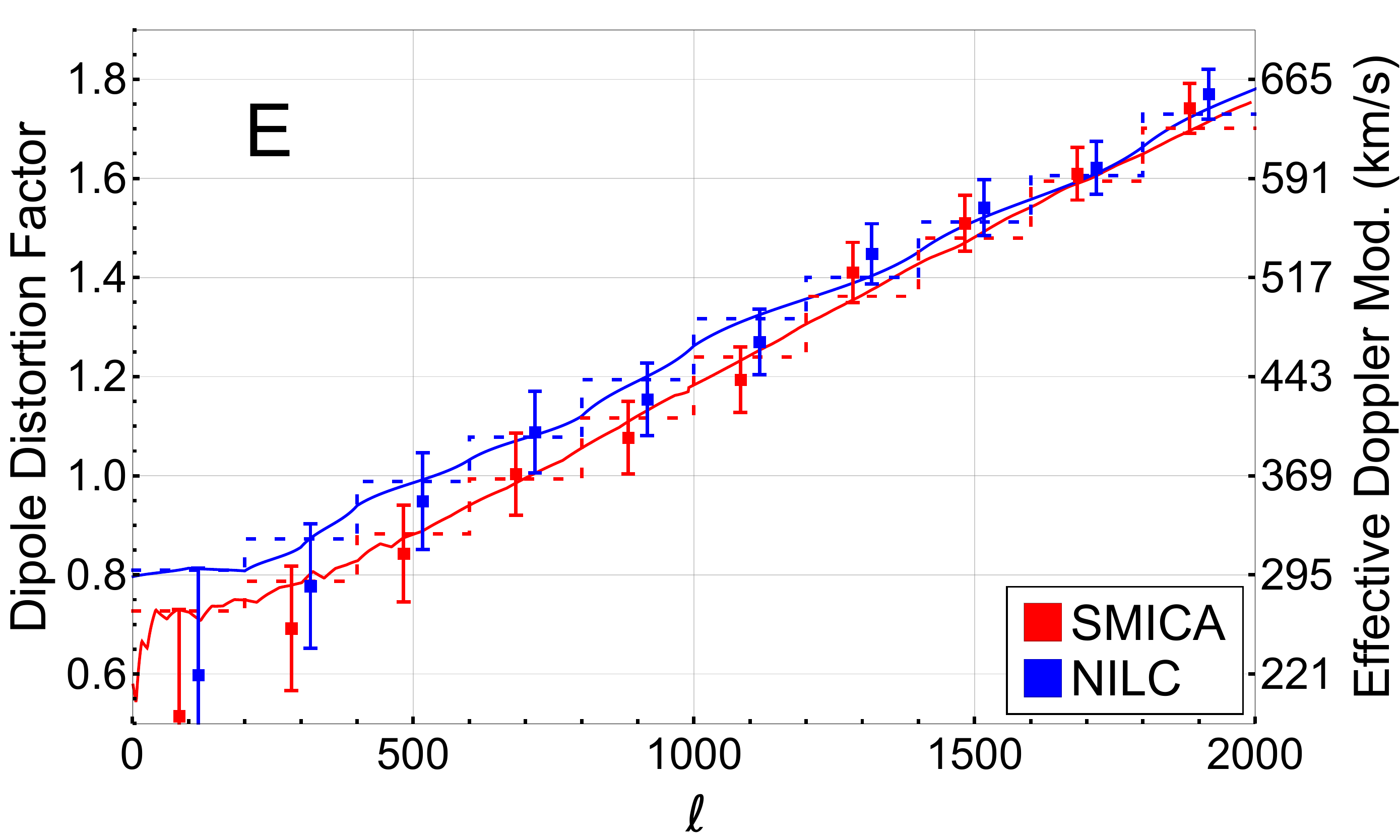}
    \caption{
    Dipole Distortion values for the $T$ and $E$ maps for both SMICA and NILC. Solid lines are computed using Eq.~\eqref{eq:dd_by_l}; dashed stepwise lines are their binned versions from Eq.~\eqref{eq:binneddd}. We also show in each bin the average recovered values in our 1024 simulations and their $1\sigma$ uncertainty as a cross-check. Using Planck's dx12 simulations yield equivalent results.
    \label{fig:dd_plot}}
\end{figure}

The value of this spurious effective velocity as a function of multipole is depicted in Figure~\ref{fig:dd_plot} using Eq.~\eqref{eq:dd_by_l}. For our pipelines (discussed below) we need to work on pixel space and thus need to make maps using bins of~$\ell$. We therefore divide the $a_{\ell m}$s into bins, generate a map for each bin separately and apply the modulation on pixel space. We settled on 10 multipole bins of size $\Delta\ell=200$ ($2\le\ell<2001$) as a good compromise between computational cost and precision: the latter was estimated to be $\sim 1\%$. The DD effect is thus the weighted average
\begin{equation}
    {\rm DD}_{\mathrm{bin}}^{\mathrm{M}}\equiv \frac{\sum\limits_{\mathrm{\ell = 200(\rm{bin}-1)+2}}\limits^{\mathrm{200\rm{bin}+1}}{(2\ell+1){\rm DD}^{\rm M}(\ell)}}{\sum\limits_{\mathrm{\ell = 200(\rm{bin}-1)+2}}\limits^{\mathrm{200\rm{bin}+1}}{(2\ell+1})} \; .
\label{eq:binneddd}
\end{equation}
The $2\ell+1$ weight is used as that is the number of modes in each $\ell$. As discussed in more detail on Appendix~\ref{app-error-by-ell} the Doppler uncertainty in each multipole is, in the absence of mask and noise, directly given by $1.225(2\ell+1)^{-1/2}$. Therefore we are weighting by the inverse uncertainty squared, so that in each bin the higher multipoles have a larger weight in the overall DD factor.

We finally define $\bm{\beta}^{\mathrm{\rm DD},\mathrm{M}}_{\rm{bin}} \equiv {\rm DD}^{\rm M}_{\rm bin}\bm{\Delta}_{1}$ and so the transformation on $T(\bm{\hat{n}})$ is for a given multipole bin is
\begin{equation}
    T(\hat{\bm{n}})^{\mathrm{DD,M}}_{\rm{bin}}=\frac{T(\hat{\bm{n}})^{\mathrm {M}}_{\rm{bin}}}{\gamma\left(1-\bm{\beta}^{\mathrm{\rm DD},\mathrm{M}}_{\rm{bin}} \cdot\hat{\bm{n}}\right)}.
\label{eq:T_dd_bin}
\end{equation}
Summing the $a_{\ell m}$s of all 10 binned maps we generate the final map with the complete DD effect, completing the DD pipeline (see Figure \ref{fig:dd_pipeline}) for both $\mathrm{M = SMICA,}\,T$ and $\mathrm{M = NILC,}\,T$.  The procedure is equivalent for $E(\hat{\bm{n}})$.

\section{Simulations Pipelines}\label{sec:simulations}

\begin{figure}
    \centering
    \includegraphics[width=.9\columnwidth]{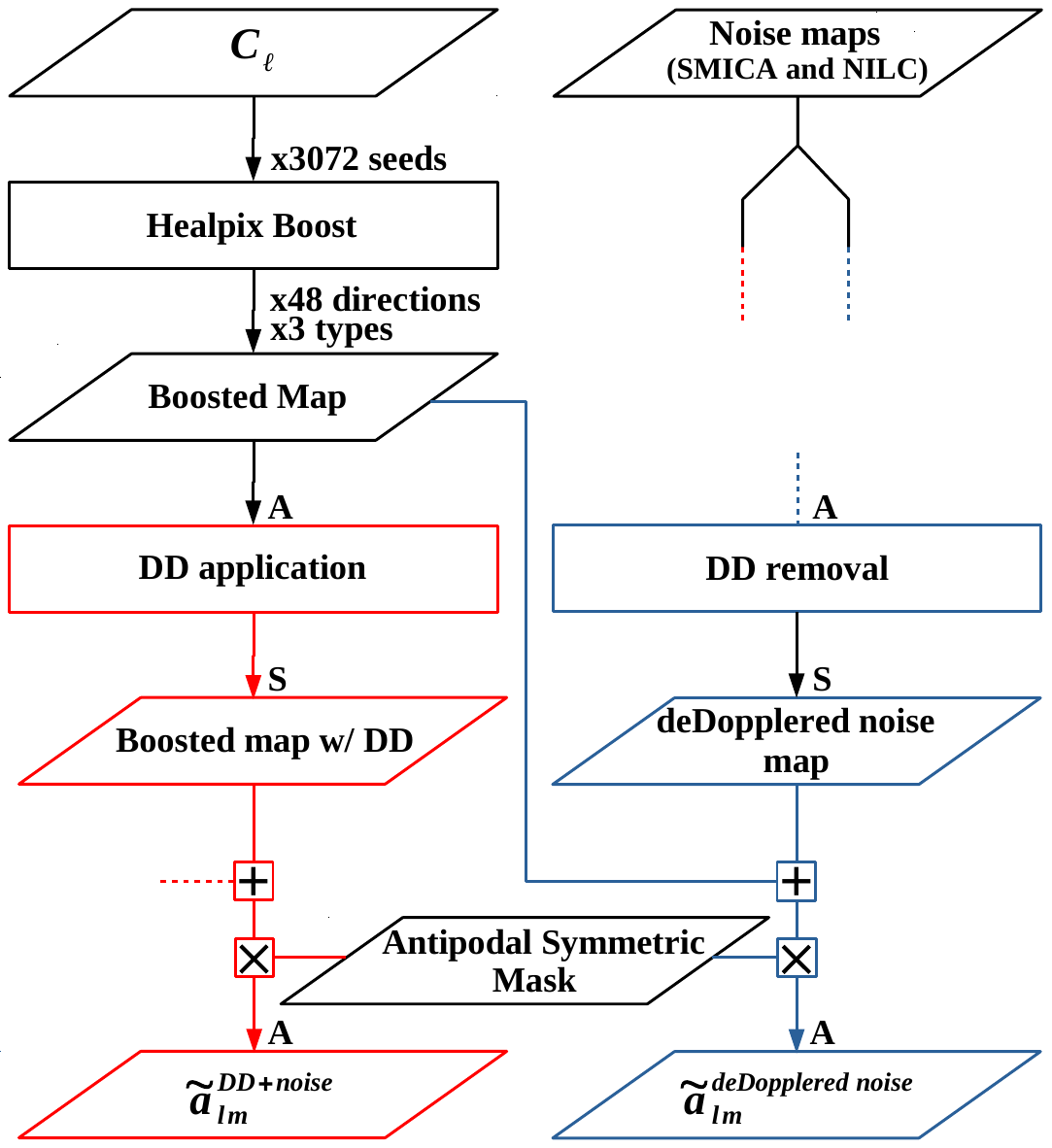}
    \caption{
    Similar to Figure~\ref{fig:dd_pipeline} for the full simulation pipelines MP and CCP. Red (blue) steps are exclusive to the MP (CCP). We generate $64$ boosted maps from Planck $C_\ell$s
    for each 48 different directions and apply one of the 3 types of signal: a complete boost (Aberration+Doppler); Doppler only; or aberration only. In total we have 9216 simulations for each SMICA and NILC. The pipelines then diverge. \textbf{MP:} The DD is applied in all maps, which are then summed with Planck noise maps and masked. \textbf{CCP:} A DD removal (dDD) is applied over the noise maps, which are then summed with the signal (A+D, D or A) and masked.
    \label{fig:simulations_pipeline}}
\end{figure}

In order to have an accurate estimate we make use of a large number of realistic simulations to fit the effective bias function depending on angular scale. Aiming at robustness, we use two alternative pipelines in order to confirm our results. The Main Pipeline (MP) is the one which we use to quote our final results. The Cross-Check Pipeline (CCP) is an alternative way to remove the DD and which we use to validate our removal method. All steps done to reproduce these simulations are summarized in Figure~\ref{fig:simulations_pipeline}.

To recreate the effects of aberration and Doppler on CMB maps we used the HEALPix-Boost code \citep{Catena:2012hq,Notari:2013iva},\footnote{\url{https://www.github.com/mquartin/healpix-boost}} a modified version of HEALPix~\citep{Gorski:2004by}. HEALPix-Boost applies both effects on the mapping between harmonic and pixel space first on the $\hat{z}$ direction and generalises to an arbitrary direction through a Wigner rotation matrix \citep[see, for instance,][]{Amendola:2010ty} using a modified \textit{alteralm} code.

The pipeline presented in Figure~\ref{fig:simulations_pipeline} can be summarized as follows. We start with a $C_{\ell}$ based on latest Planck 2018 cosmology~\citep{Aghanim:2018eyx} including lensing. We then produce $3072$ Gaussian maps with HEALPix. For each map 3 types of signal are applied using HEALPix-Boost, generating 3 new maps: a complete boost (Aberration + Doppler), a Doppler-only effect and an aberration-only effect. In the most  general case, Doppler and aberration may have different orientations, so we simulate each of the 3 effects in 48 different directions (based on the center of HEALPix pixels of $\texttt{Nside}=2$).  In each direction and signal type we perform 64 simulations. In all cases we assume $|\bm{\beta}|=0.001234$ (i.e. $v \equiv \beta c = 370$ km/s). Note that since all 3 cartesian estimators are independent, this procedure is equivalent to making boosts in each component for different values in the range $[-0.001234, 0.001234]$. We remark that this procedure differs from what was carried out by~\cite{Aghanim:2013suk}, where all boosted simulations were on the dipole direction with the dipole amplitude. The motivation behind our procedure is to guarantee that both aberration and Doppler estimators are not biased towards the dipole direction.

After this step there is a divergence between the Main Pipelines (MP) and cross-check pipeline (CCP). The main difference between the MP and CCP is that on MP the DD effect is left to be removed later in the analysis while on the CCP the DD is removed directly on the CMB maps. Since we compute mask and estimator biases directly using simulations each pipeline lead to different biases to be removed on the analysis. The base code to reproduce the MP and CCP, including examples, is made available online.\footnote{\url{https://www.github.com/pdsferreira/cmb-aber-dopp}}

\subsection{MAIN PIPELINE (MP)}

\begin{figure}
\centering
    \includegraphics[width=\columnwidth]{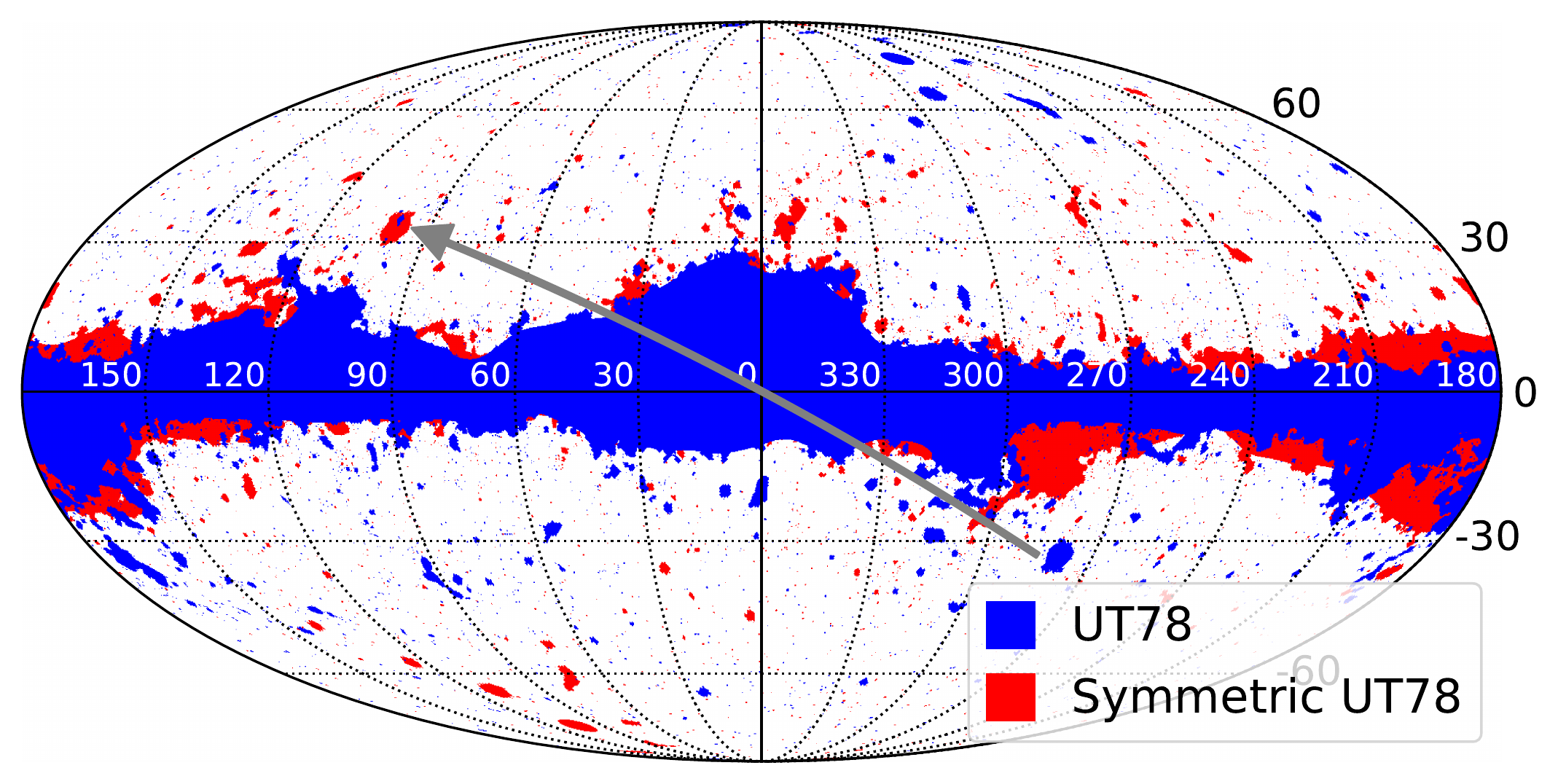}
    \caption{Antipodally symmetric UT78(2018) mask. The grey arrow connects the Large Magellanic Cloud original position and its antipodal position. This removes an extra $5\%$ of sky but helps preventing contamination of the mask in the estimators.
    \label{fig:symm_mask}}
\end{figure}

For the MP the DD effect is applied over the simulated maps (already with aberration, Doppler or boost signals) and we arrive at $T(\hat{\bm{n}})^{\rm{DD,M}}$ in Eq.~\eqref{eq:T_dd_bin}. The maps are then summed with the Planck noise maps $N^{\rm{M}}(\bm{\hat{n}})$ for each component separation method $\rm{M}$ and multiplied by the mask $W(\bm{\hat{n}})$. We will henceforth use a tilde to denote variables which include not only the masking effect but also the anisotropic noise. The final, realistic, map is
\begin{equation}
    \tilde{T}(\hat{\bm{n}})^{\mathrm{DD+noise,M}}=
    \left[T(\hat{\bm{n}})^{\mathrm{DD,M}}+N^{\rm{M}}(\hat{\bm{n}})
    \right]W(\hat{\bm{n}}) \, .
\end{equation}
and equivalently for $E$ maps. In order to avoid any dipolar signature from the mask, which could be a source of bias, we use a mask with antipodal symmetry, as proposed in~\cite{Quartin:2014yaa}. This type of mask has the special property
\begin{equation}
    W(l ,b)=W(l +\pi,-b) \, ,
    \label{eq:antipodal_symm}
\end{equation}
where $(l,b)$ are the galactic coordinates. So for any masked pixel we mask also the antipodal pixel, ensuring by construction a dipolar symmetry. We generate our masks based on the component separation common masks for both temperature (also known as UT78(2018) mask) and polarization (also known as UP78(2018) mask), apodized with a $10'$ Gaussian beam. This process removes an extra 5\% of the sky, which diminishes the precision of the estimators by only $\sim 2\%$. Figure~\ref{fig:symm_mask} depicts both the original and the antipodally symmetric version of Planck UT78(2018) mask, which we employ in this work for the $T$ maps.

Finally we compute the $a_{\ell m}$s of the final map. We use these simulations on the statistical and bias analyses. All the steps are carried out using $\texttt{Nside}=2048$ maps (i.e. $\sim50$ \rm{megapixels}), $\ell_{\mathrm{max}}=2001$ and considering the realistic beam function given for each of Planck component separation maps.

\subsection{CROSS-CHECK PIPELINE (CCP)}
In the CCP instead of adding the DD to the simulations, we remove it from the final Planck maps by applying a negative Doppler boost. This DD removal produce ``de-Dopplered'' (dDD) CMB maps. To be consistent, we also remove the DD from the noise maps, again with a negative Doppler boost. The (dDD) noise maps are calculated in bins, as done in Eq.~\eqref{eq:T_dd_bin}:
\begin{equation}
    N(\hat{\bm{n}})^{\mathrm{dDD,M}}_{\rm bin}= \frac{N(\hat{\bm{n}})^{\mathrm{dDD,M}}_{\rm bin}}{\gamma\left(1+2{\rm DD}^{\rm M}_{\rm bin} \bm{\Delta}_{1} \cdot \hat{\bm{n}}\right)} \,.
\end{equation}
The final noise summed maps are:
\begin{equation}
    N(\hat{\bm{n}})^{\mathrm{dDD,M}} = \sum\limits_{\text{bin=1}}\limits^{10} N(\hat{\bm{n}})^{\mathrm{dDD,M}}_{\rm bin} \, .
\end{equation}
$N(\hat{\bm{n}})^{\mathrm{dDD,M}}$ is then summed with the maps containing the signal (aberration, Doppler or boost) and the result multiplied by the mask:
\begin{equation}
    \tilde{T}(\hat{\bm{n}})^{\mathrm{dDD,M}}= \left[T(\hat{\bm{n}})+N(\hat{\bm{n}})^{\mathrm{dDD\,noise,M}} \right]W(\hat{\bm{n}}) \, ,
\end{equation}
for $T$ maps and similarly for $E$ maps (using polarization noise maps). From this we compute the $a_{\ell m}s$, completing the CCP for simulations.

This can be done equivalently on harmonic space by subtracting the DD terms directly on the $a_{\ell m}s$. From Eq.~\eqref{eq:L-2nd-order}, considering $\ell>2$ terms and using the definition of Eq.~\eqref{eq:dd_by_l} the DD can be removed using:
\begin{equation}
    a^{\mathrm{dDD,M}}_{\ell m} = a^{\mathrm{M}}_{\ell m}- \left[T(\bm{\hat{n}^\prime})\times(\Delta_1(\bm{\hat{n}^\prime}) \cdot \bm{\hat{n}^\prime})\right]_{\ell m}\mathrm{DD^{M}_{\ell}} \; ,
\end{equation}
for both $T$ and $E$. The term $\mathrm{DD^{M}_{\ell}}$ assumes the same value for all $m$ modes of a given $\ell$. After that the mask should be applied on the map generated by $a^{\mathrm{dDD,M}}_{\ell m}$. As we are applying  on the last term an additional DD on the observed maps (which already include a DD contribution), this ignores quadratic effects.  This process also removes the DD from the noise, to be consistent one also have to remove the DD from the noise maps when estimating the bias:
\begin{equation}
    a^{\mathrm{dDD,noise,M}}_{\ell m}\!= a^{\mathrm{noise,M}}_{\ell m}- \left[N(\bm{\hat{n}^\prime})\!\times\!(\Delta_1(\bm{\hat{n}^\prime})\!\cdot\!\bm{\hat{n}^\prime})\right]_{\ell m}\mathrm{DD^{M}_{\ell}} .
\end{equation}
Both map and harmonic based dDD methods above recover the same results for $TT$ estimator and very similar results on $EE$. In our companion Letter~\cite{Ferreira:2020aqa} we make use of the map based method in our final results with the CCP on real data.

In both cases above of the CCP besides the need to adjust the noise maps
some of the statistical properties of the maps may be affected. Indeed this was not considered in Planck's mapmaking procedures or noise simulations. In particular single-frequency weights used in the component separation may no longer be the optimal ones, due to the DD removal. This is the reason we do not set this as the main pipeline. We nevertheless find that both pipelines give consistent results, which increases the robustness of our findings.

\section{Bias removal through brute-force mock simulations}\label{sec:bias}
Besides the DD, other effects need to be accounted for in order to produce unbiased estimations for $\bm{\beta}^{\rm A}$ and $\bm{\beta}^{\rm D}$. In particular we need to consider the anisotropy of the noise, the effect of masking and the correlation between $\bm{\beta}^{\rm A}$ and $\bm{\beta}^{\rm D}$. On the MP we minimize the influence of DD, noise and mask by applying a sum and multiplication bias correction on the results of the estimator for each multipole bin of $\Delta \ell = 10$, for each case (aberration, Doppler or boost), cartesian component and mapmaking method.

\begin{figure*}
    \includegraphics[width=1.89\columnwidth]{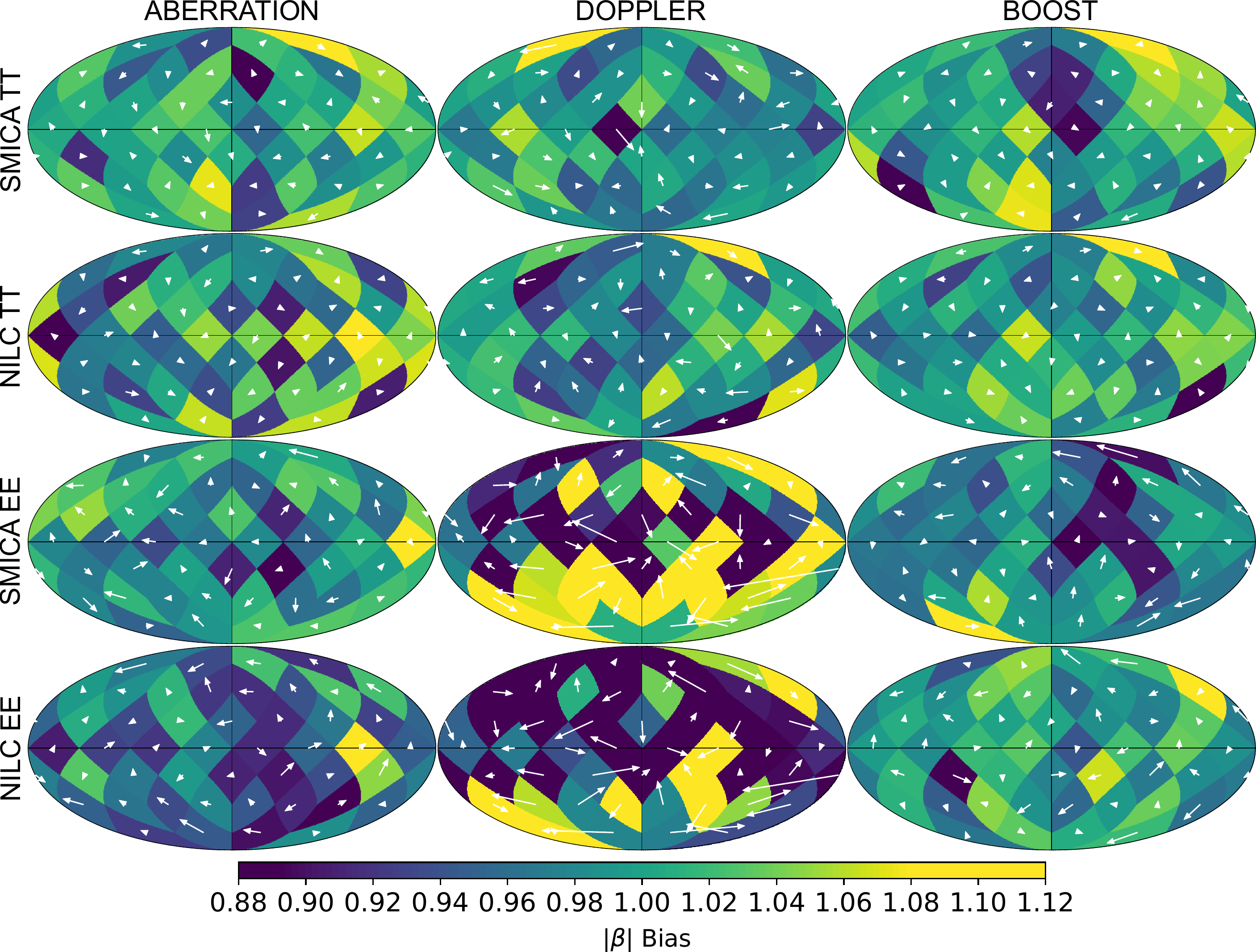}
    \caption{Residual bias maps after bias removal for simulations with noise, mask and the DD effect, for the 3 estimators, component separation method and for $TT$ and $EE$ correlations. White arrows start on the simulated $\bm{\beta}$ directions and end on the recovered ones. Color indicates the $|\bm{\beta}|$ bias factor. Results in each of the 48 directions are the average over 64 simulations with $\ell_{\mathrm{max}}=1800$ (1150) for $TT$ ($EE$).
    \label{fig:bias_maps}}
\end{figure*}

This bias correction is estimated allowing for a signal on any of 48 sky directions, and any $48^2$ combinations of $\bm{\beta}^{\rm A}$ and $\bm{\beta}^{\rm D}$. The bias is then fitted for each cartesian component estimator separately. As discussed in Section~\ref{sec:simulations}, this means that for each component we allow a large range of values of aberration and/or Doppler. The bias correction should thus work for any combination of such components.

In practice, we find the sum and multiplication bias correction factors by minimizing the following $\chi^2$:
\begin{equation}
    \chi^2 = \sum_{n} \left(
    \frac{\tilde{\beta}_{nXYi,{\rm bin}}^{\mathrm{SIM,M}}-\lambda_{Xi,{\rm bin}}^{\rm M} \beta_{nXYi}^{\mathrm{FID}} - \mu_{Xi,{\rm bin}}^{\rm M}}{\tilde{\sigma}_{XYi,{\rm bin}}^{\mathrm{SIM,M}}}    \right)^{2},
\end{equation}
where $n$ is a label for each simulation that identifies the original Gaussian simulation seed and the direction of the (A, D or B) signal, $X$ is the signal applied (A, D or B), $Y$ is the estimator used (A, D or B), $i$ is the cartesian component and ``bin'' is one of 200 $\ell$-bins between $\ell=2$ and $\ell=2001$. $\rm{SIM}$ means the result of the estimator over a simulation, $\rm{FID}$ is the fiducial value and $\rm{M}$ is the mapmaking method (SMICA or NILC). $\lambda_{Xi,{\rm bin}}$ is the multiplicative bias and $\mu_{Xi,{\rm bin}}$ the sum bias. Our cartesian coordinate system is related to the galactic coordinate system as follows. $\hat{x}$: $ (\ell,b) \rightarrow (0,0)^{\circ}$, $\hat{y}$: $ (\ell,b) \rightarrow (90,0)^{\circ}$, $\hat{z}$: $(\ell,b) \rightarrow (0,90)^{\circ}$.

The estimated bias must not depend on the signal applied, and only on the estimator used, as we do not know \emph{a priori} the signal. The bias coefficients $\lambda_{Xi,{\rm bin}}$ and $\mu_{Xi,{\rm bin}}$ therefore cannot depend on $\bm{\beta}$. We use $X=Y$ to fit $\lambda_{Xi,{\rm bin}}$ and $\mu_{Xi,{\rm bin}}$ and $X \ne Y$ to fit the correlations factors $R^{\,}_{i}$ and $S^{\,}_{i}$ (see Section~\ref{ap:correlation}). After removing the correlation the results are correct for any combinations of $X$, $Y$. The unbiased results with $X=Y$ and $X \ne Y$ are used to calculate the statistical and systematical errors, respectively. The final unbiased results are given by
\begin{equation}
    \beta_{nXYi,{\rm bin}}^{\mathrm{SIM,M}}= \frac{\tilde{\beta}_{nXYi,{\rm bin}}^{\mathrm{SIM,M}}-\mu_{Xi,{\rm bin}}^{\rm M}}{\lambda_{Xi,{\rm bin}}^{\rm M}} \, .
\end{equation}
The bins can be summed to find the vector result
\begin{equation}
    \beta_{nXYi}^{\mathrm{SIM,M}}\!\!=\!\!\left[\sum\limits_{\rm bin=1}^{\mathrm{maxbin}}\!\!\frac{\beta_{nXYi,bin}^{\mathrm{SIM,M}}}{{\sigma^{\rm{SIM,M}}}_{XYi,{\rm bin}}^{2}}\right]\!\!\! \left[\sum\limits_{\rm bin=1}^{\mathrm{maxbin}}\!\! \frac{1}{{\sigma^{\rm{SIM,M}}}_{XYi,{\rm bin}}^{2}}\right]^{\!-1} \!\!\! ,
\end{equation}
so that $\bm{\beta}_{nXY}^{\,\mathrm{SIM,M}}= \{\beta_{nXYx}^{\mathrm{SIM,M}}, \beta_{nXYy}^{\mathrm{SIM,M}},\beta_{nXYz}^{\mathrm{SIM,M}}\}$. The value of the bias coefficients and maxbin depends on whether we are using the $TT$ or $EE$ correlations. Based on the expected uncertainties shown in Section~\ref{app:ap_ideal_stat}, we note that there little information in $TT$ ($EE$) for $\ell>1800$ ($\ell>1150$), so we set these value as our maximum multipole in the estimators. In what follows we will always remove the first DD bin  ($2<\ell<201$). Using $\ell_{\rm min} = 202$  avoids any influence from large-scale anomalies in the real data and the signal-to-noise ratio in this first bin is minimal, as shown in Appendix \ref{app-error-by-ell}.

The estimated cartesian components are the final results of the pipeline. It is also useful to quote results in galactic spherical coordinates. This however introduce a third bias, as the best fit estimators for each component of a vector does not yield directly the best estimator of its amplitude. In this case we also renormalize the amplitude of the vectors dividing by a factor $\nu_{Xi,{\rm bin}}^{M}$, the ``amplitude renormalization''. This ensures that, after correlation removal, $\langle| \bm{\beta}_{XY}^{\,\mathrm{SIM,M}}|\rangle = |\beta_{XY}^{\mathrm{FID}}|$. This correction is a general property of recovering the amplitude of vectors from their components, and $\nu_{Xi,{\rm bin}}^{M}$ is always larger than 1. Therefore, the uncertainties in the amplitude are smaller than the average error in the components.

In Figure~\ref{fig:bias_maps} the residual bias maps are shown. These maps represent the difference between fiducial and recovered results by all estimators, after bias and correlation removal, on simulations for signals applied on different directions. Each arrow starts on the expected direction of the simulated signal and ends on the recovered direction of $\bm{\beta}$, using the average of unbiased results of 64 simulations. The color indicates the $|\bm{\beta}|$ bias factor.
This figure indicates that in most cases the systematics are negligible when compared with the statistical uncertainties. The only exceptions are the Doppler $EE$ measurement, which have significant residual biases in both SMICA and NILC. However, Doppler $EE$ has low expected $S/N$ and this therefore does not affect significantly the final Doppler $TT$+$EE$ results in our companion paper~\cite{Ferreira:2020aqa}.
Table~\ref{tab:errors_stat_syst} reports the comparison of statistical and systematic uncertainties in all cases for the absolute values and directions. We remark that although around only 1 in 8 of the directions have significant residual bias on longitude, most have statistically-significant residual bias on the absolute value.

\setlength\tabcolsep{2pt}
\begin{table}
    \centering
    \begin{tabular}{ c l c c c c c c }
    \cmidrule{1-8}\morecmidrules\cmidrule{1-8}
    & &  \multicolumn{2}{c}{$|\bm{\beta}| \; \mathrm{(km/s)}$}  & \multicolumn{2}{c}{$l(^{\circ})$} & \multicolumn{2}{c}{$b(^{\circ})$} \\
    \cmidrule{1-8}\morecmidrules\cmidrule{1-8}
    $\boldsymbol{TT}$  & estimator  &   stat. & syst.$\;$   & $\;$stat. & syst.$\;$   & $\;$stat. & syst.       \\
    \hline
    & Aberration \rule{0pt}{3ex}    & 130  & 13 & 36 & 1  & 21 & 1 \\
    \raisebox{0pt}[0pt][0pt]{\rotatebox[origin=c]{90}{SMICA}} & Doppler     & 150  & 13  & 57 & 3  & 31 & 1\\
    & Boost  \rule[-1.5ex]{0pt}{0pt} & 100  & 11  & 23  & 0.1  & 15 & 0.2 \\
    \hline
    & Aberration \rule{0pt}{3ex}     & 130  & 13 & 35  & 2  & 22 & 0.3 \\
    \raisebox{-5pt}[0pt][0pt]{\rotatebox[origin=c]{90}{$\;\;\;$ NILC}} & Doppler        & 150  & 10  & 60  & 3  & 30 & 1  \\
    & Boost   & 100  &  11 & 22 & 1  & 15 & 0.1 \\
    \cmidrule{1-8}\morecmidrules\cmidrule{1-8}
    $\boldsymbol{EE}$   & estimator  &   stat. & syst.$\;$   & $\;$stat. & syst.$\;$   & $\;$stat. & syst.       \\
    \hline
    & Aberration \rule{0pt}{3ex}   & 160  & 17  & 66 & 0.1  & 28 & 1  \\
    \raisebox{-6pt}[0pt][0pt]{\rotatebox[origin=c]{90}{$\;\;\;$ SMICA}} & Doppler    & 340 & 380 & 103  & 6 & 34 & 1 \\
    & Boost \rule[-1.5ex]{0pt}{0pt}  & 150  & 14  & 64  & 3  & 29 & 0.5  \\
    \hline
    & Aberration  \rule{0pt}{3ex}   & 150 & 15 & 65  & 0.3  & 29 & 1  \\
    \raisebox{-5pt}[0pt][0pt]{\rotatebox[origin=c]{90}{$\;\;\;$ NILC}} & Doppler    & 330 & 350  & 100  & 5  & 35 & 1  \\
    & Boost  & 150 & 15 & 61  & 1  & 29 & 0.2 \\
    \cmidrule{1-8}\morecmidrules\cmidrule{1-8}
    $\boldsymbol{TT}$+$\boldsymbol{EE}$  & estimator  &   stat. & syst.$\;$   & $\;$stat. & syst.$\;$   & $\;$stat. & syst.       \\
    \hline
    & Aberration  \rule{0pt}{3ex}    & 99  & 13  & 32  & 0.1  & 19 & 0.7 \\
    \raisebox{1pt}[0pt][0pt]{\rotatebox[origin=c]{90}{SMICA}} & Doppler        & 140 & 13 & 56 & 3 & 28 & 0.5  \\
    & Boost  & 84 & 9 & 21  & 0.1  & 14 & 0.2 \\
    \hline
    & Aberration \rule{0pt}{3ex}     & 100 & 10 & 32  & 0.3  & 20 & 0.3 \\
    \raisebox{-0.5pt}[0pt][0pt]{\rotatebox[origin=c]{90}{NILC}} & Doppler        & 140  & 10  & 56  & 2 & 28 & 0.2 \\
    & Boost  & 83  & 9  &  22 & 1  & 15 & 0.1 \\          \cmidrule{1-8}\morecmidrules\cmidrule{1-8}
    \end{tabular}
    \caption{Statistical and systematic uncertainties in galactic coordinates for each component separation method and   each estimator for the Main Pipeline simulations. Since the uncertainty on $l$ depends on $b$ we present their average values.}
    \label{tab:errors_stat_syst}
\end{table}

\section{Removing the leakage between aberration and Doppler}\label{ap:correlation}

Since both aberration and Doppler introduce $\ell, \ell+1$ correlations in harmonic space, their estimators have an inherent correlation. To remove this correlation we define 2 coefficients of signal leakage, dubbed $R^{\,}_{i}$ (Doppler to aberration) and $S^{\,}_{i}$ (aberration to Doppler), for each cartesian component $i$. We start by assuming that the correlated results $\beta^{[c]}_{X,i}$ are related to the uncorrelated estimators as:
\begin{align}
    \beta_{A,i} \equiv \beta_{A,i}^{\mathrm{[c]}}-R^{\,}_{i}\beta_{D,i} \; ,\label{eq:correlation_approx_a} \\
    \beta_{D,i} \equiv \beta_{D,i}^{\mathrm{[c]}}-S^{\,}_{i}\beta_{A,i} \; \label{eq:correlation_approx_d}.
\end{align}

To fit $S^{\,}_{i}$ we minimize the following $\chi^2$:
\begin{align}
    \chi_S^2 \,=\; & \sum_{n}\left(\frac{\beta^{\mathrm{[c]}}_{n A \, D i}-\beta^{\mathrm{FID}}_{n A \, A i}S_{i}}{\sigma_{D i}} \right)^2 \nonumber  \\
    &+ \sum_{n}\left(\frac{\beta^{\mathrm{[c]}}_{n B \, D i}-\beta^{\mathrm{FID}}_{n B \, B i}(S_{i}+1)}{\sigma_{D i}} \right)^2 \; ,
    \label{eq:correlation_factorAD}
\end{align}
where $n$ is the simulation id, the second and third sub-indices are the simulation signal and estimator used and $\sigma_{X i}$ the standard deviation of the estimator. The first sum makes use only of simulations where the expected result is null if there is no correlation ($S_i=0$). On the second sum, which is for the boost case, the Doppler estimation should recover the correct velocity, and hence the factor $(S_{i}+1)$. Equivalently for $R^{\,}_{i}$ we have
\begin{align}
    \chi_R^2 \,=\; & \sum_{n}\left(\frac{\beta^{\mathrm{[c]}}_{n D \, A i}-\beta^{\mathrm{FID}}_{n D \, D i}R_{i}}{\sigma_{A i}} \right)^2 \nonumber \\
    &+ \sum_{n}\left(\frac{\beta^{\mathrm{[c]}}_{n B \, A i}-\beta^{\mathrm{FID}}_{n B \, B i}(R_{i}+1)}{\sigma_{A i}} \right)^2 \,.
    \label{eq:correlation_factorDA}
\end{align}
Using~\eqref{eq:correlation_approx_a} and \eqref{eq:correlation_approx_d} we then get the approximation
\begin{equation}
    \beta_{A,i} \simeq \frac{\beta_{A,i}^{\mathrm{[c]}}-R^{\,}_{i}\beta_{D,i}^{\mathrm{[c]}}}{1-R^{\,}_{i}S^{\,}_{i}} \,,\qquad \beta_{D,i} \simeq \frac{\beta_{D,i}^{\mathrm{[c]}}-S^{\,}_{i}\beta_{A,i}^{\mathrm{[c]}}}{1-S^{\,}_{i}R^{\,}_{i}}\,. \label{eq:correlation_removal}
\end{equation}

\begin{figure}[t]
    \centering
    \includegraphics[ width=0.9\columnwidth]{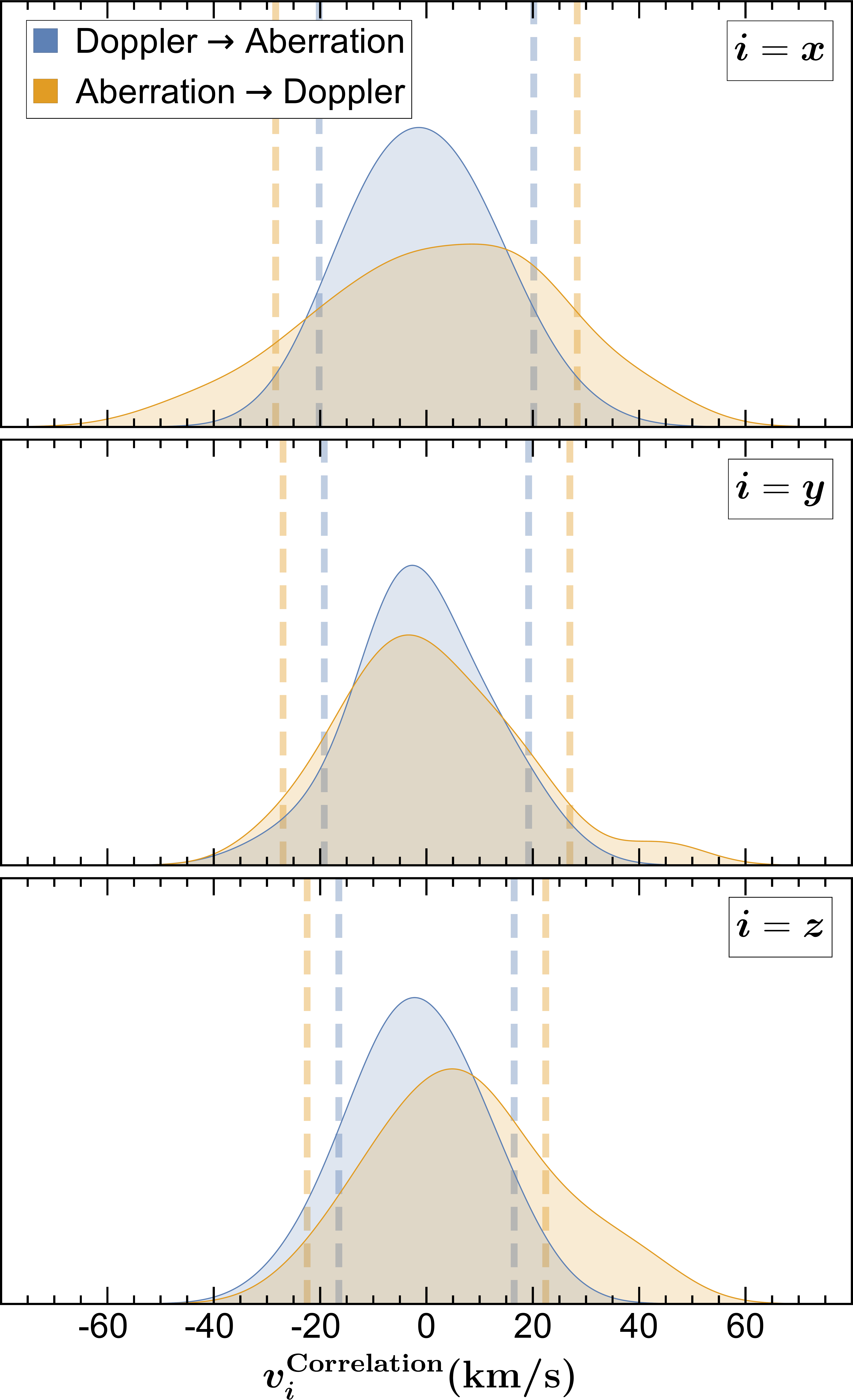}
    \caption{Average residual leakage between the $TT$+$EE$ estimators for each cartesian component. Doppler $\to$ Aberration represents the aberration estimator results for a simulated Doppler modulation of $v_{D}=370$ km/s without any aberration, and \emph{vice-versa}. Dashed lines mark the ideal $1\sigma$ (standard error) statistical intervals. We conclude that residual leakages are small and can be neglected.
    \label{fig:xyz_correlation}
    }
\end{figure}

\begin{figure*}
    \includegraphics[ clip, width=1.83\columnwidth]{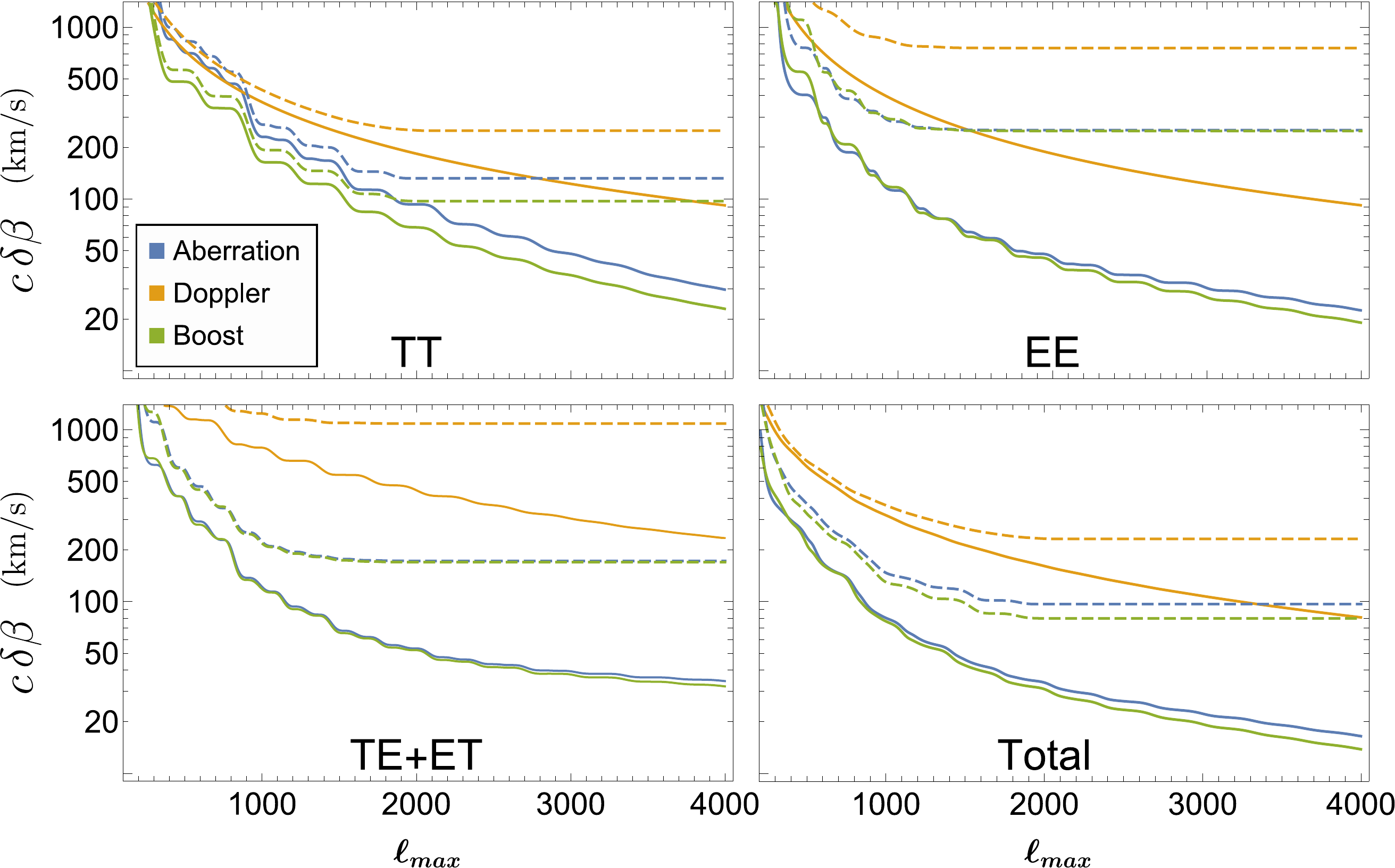}
    \caption{Expected uncertainties in each estimator as a function of $\ell_{\rm max}$ for Planck (2018). Dashed lines represents the Planck $1\sigma$ errors, solid lines the ideal theoretical calculations without noise and mask.
    \label{fig:planck-forecasts}}
\end{figure*}

The effectiveness of this method can be seen comparing the $\beta_{A,i}$ and $\beta_{D,i}$ obtained using this method with the ideal (simulated) case, which we depict in Figure~\ref{fig:xyz_correlation}. As can be seen the residual leakage are small and can be neglected, and our approach is sufficient for the precision we can achieve. This is done separately for $TT$ and $EE$ measurements.

\section{Forecast uncertainties with Planck and future CMB experiments}\label{app:ap_ideal_stat}

We start by comparing the expected theoretical uncertainties in each estimator with the simulated uncertainties for the $TT$, $TE$+$ET$ and $EE$ cases for the case of Planck. These estimates follow the approach in~\cite{Amendola:2010ty}, but generalized to separate the Doppler, aberration and boost cases. In Figure~\ref{fig:planck-forecasts} we depict both the ideal errors without noise and mask, and the Planck 2018 statistical errors using the symmetric masks ($f_{\rm sky}= 0.73$) for temperature and polarization, $N_{\ell}$ based on the dx12  SMICA Planck simulations and the realistic effective beam. 
As the mask affects each direction differently, we show the average of the cartesian components estimators as a function of $\ell_{\rm max}$.

As can be seen, the $EE$ correlations in Planck can provide a measurement of aberration or boost at a around $1.5\sigma$, while for Doppler the errors are much larger. This is due to the low S/N in the $a_{\ell m}^{\rm E}$; with less noise the $EE$ could provide a similar precision to $TT$ -- see~\cite{Notari:2011sb,Burigana:2017bxl}. Note also that for $EE$ (and with less significance also for $TE$) a partial cancellation between Doppler and aberration makes the boost estimator less precise than the pure aberration one for some ranges of $\ell_{\rm max}$. This is explained in detail in Appendix~\ref{app-error-by-ell} where we explore the uncertainties in each estimator as a function of $\ell$. The actual measured uncertainties in the real data, which include systematic effects and the leakage between the Doppler and aberration signals is similar (see below), and the results using Planck 2018 data are discussed in detail in the companion letter~\cite{Ferreira:2020aqa}.

As discussed in Section~\ref{sec:bias}, an unbiased estimate of the vector amplitude $|\boldsymbol{v}|$ requires an amplitude normalization factor $\nu_X^M$. For Planck, this factor  ranges from around 1.0 to around 1.8, depending on the estimator ($\beta^{\rm A}$, $\beta^{\rm D}$ and $\beta^{\rm B}$), $\ell_{\rm max}$, component separation method (SMICA or NILC) and whether we are using $TT$ or $EE$ correlations.

As can be seen from Figure~\ref{fig:planck-forecasts}, for Planck adding $EE$ may improve slightly the $TT$ precision. The $TE$ and $ET$ cross-correlations could also be used to provide extra precision to aberration. The analysis is however more involved for them for 3 reasons: $(i)$ the DD effect is not straightforward; $(ii)$ our simple scheme to remove leakages between signals cannot be used as is; and $(iii)$ our estimators would become correlated~\citep[see][]{Amendola:2010ty}.  These difficulties are ignored in Figure~\ref{fig:planck-forecasts}.  Since the expected improvement in precision would be small, specially on Doppler which has the largest uncertainty in~\cite{Ferreira:2020aqa} we used only $TT$+$EE$ in our real data analysis.

We also provide forecasts for two upcoming ground-based CMB experiments: the Simons Observatory and the CMB Stage IV experiment (CMB-S4). For the Simons Observatory we use the specifications of the optimistic case available in~\cite{Ade:2018sbj}. For the CMB-S4 we use the precision goal of the experiment discussed in~\cite{Abazajian:2016yjj}, which assume $\sigma_T = 1\,\mu K\,$arcmin and a beam of $1^\prime$ full-width half-maximum. For both experiments we consider $f_{\rm sky}=0.4$. For CMB-S4 we assume for simplicity $\ell_{\rm max} = 4000$, which yields similar results to the assumption in~\cite{Abazajian:2016yjj} that for $T$ ($E$) foregrounds can be suppressed until $\ell_{\rm max} = 3000$ (5000). These forecasts assume that, similar to Planck, all biases from masking and from the correlation between aberration and Doppler signals can be suppressed to an amount negligible compared to the statistical errors. They also treat the effects of masking and beaming in a simpler way, to wit following Eqs.~(21) and (22) of~\cite{Notari:2011sb}.

\begin{figure*}
    \includegraphics[clip, width=1.83\columnwidth]{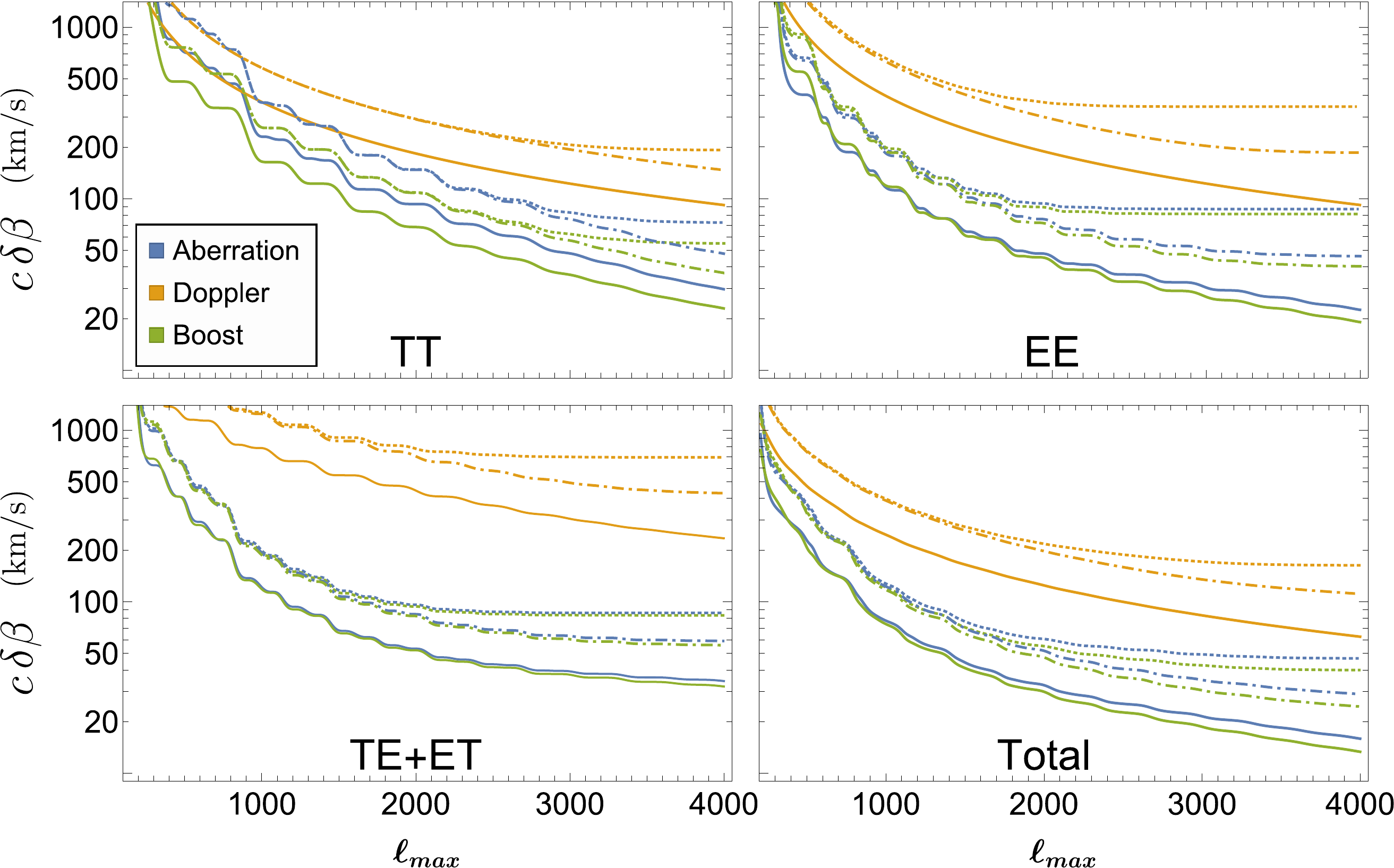}
    \caption{Same as Figure~\ref{fig:planck-forecasts} for the Simons Observatory and CMB-S4 forecasts. We assume $f_{\rm sky}=0.4$. Dotted lines represents the $1\sigma$ errors for Simons Observatory, dot-dashed lines CMB-S4 and solid lines the ideal case without noise or mask.
    \label{fig:future-forecasts}}
\end{figure*}

Figure~\ref{fig:future-forecasts} depict these forecasts for $TT$, $TE$+$ET$, $EE$, and the combination of all signals as a function of $\ell_{\rm max}$. Table~\ref{tab:errors} contains the final numbers including all multipoles. As can be seen, the precision in Doppler will remain much smaller than the one in aberration, and the difference between both will become even larger. Not even CMB-S4 should measure Doppler with a $4\sigma$ confidence level in the standard scenario in which the intrinsic dipole is negligible.

\section{Discussion}

We propose here two pipelines to estimate Doppler and aberration from CMB data. They make use of an idealist estimator as a baseline and many simulations which include realistic noise simulations and masks to remove the biases they introduce. We also shown how the Dipole Distortions, which produce a spurious Doppler signal which is degenerate with the dipole, can be removed by carefully estimating the contribution of each frequency in the each of the CMB maps. Both pipelines here were shown to be able to provide measurements of Doppler and aberration which are approximately independent among themselves and with the temperature dipole. The Main Pipeline is able to measure both signals in an unbiased way (i.e. with minimal systematic effects) with a precision very similar to the theoretical expectations: we find a $15-25\%$ larger uncertainty depending on the estimator and signal type after bias and signal leakage removal. The proposed Cross-Check Pipeline is able to produce very similar results and therefore be used to provide more robustness to the measurements.

As discussed in Section~\ref{sec:simulations}, using the non-symmetric mask the uncertainties are expected to reduce by $\sim 2\%$, but the behavior of the bias becomes much more dependent on the direction of the aberration and Doppler signal. To solve this, one should use more simulations including more directions and possibly an iterative bias removal solution, to obtain the same systematic errors, instead of fit a bias independent of direction as done here. We estimate that to get good results the computational cost will become at least 4 times higher. Since the symmetric mask removes by construction any dipolar asymmetry on signal that could possibly bias the results it improves robustness. We thus consider the symmetric mask the preferred solution for Planck. But for future ground-based experiments which cover less than half of the sky one cannot symmetrize the mask, and our bias removal pipeline will require more simulations. Apart from a higher computational cost, we expect that our proposed pipeline should also work for them with only minor modifications.

\setlength\tabcolsep{4pt}
\begin{table}
    \centering
    \begin{tabular}{ l c c c}
    \cmidrule{1-4}\morecmidrules\cmidrule{1-4}
    $c\,\delta\beta$ (km/s)& \multicolumn{3}{c}{$TT+TE+ET+EE$}  \\
    \cmidrule{1-4}\morecmidrules\cmidrule{1-4}
     & Aberration & Doppler & Boost       \\ \hline
    Planck 2018  \rule{0pt}{3ex}        & 97 & 230 & 80  \\
    Simons Observatory      & 47 & 163 & 40  \\
    CMB-S4                  & 29 & 111 & 25  \\
    Ideal $(\ell_{\rm max}\!=\!2000)$  & 33 & 124 & 30 \\
    Ideal $(\ell_{\rm max}\!=\!3000)$  & 22 & 83 & 19 \\
    Ideal $(\ell_{\rm max}\!=\!4000)$  & 16 & 63 & 13 \\
    \cmidrule{1-4}\morecmidrules\cmidrule{1-4}
    \end{tabular}
    \caption{Forecast final statistical error for future experiments in a given cartesian component. Since the mask affects each component differently, this is the expected average uncertainty in all 3 components. These estimates do not include possible residual signal leakage between Doppler and aberration. The uncertainty in $|\boldsymbol{v}|$ will be smaller (see Section~\ref{sec:bias}). With our proposed pipeline we find $\simeq20\%$ larger uncertainties after bias and signal leakage removal.
    }
    \label{tab:errors}
\end{table}

The independent measurements of aberration and Doppler opens up a new window into the early universe. They are measured with a joint significance of over $4\sigma$ in Planck 2018 and, when combined with the dipole, provide the first constraints in the CMB intrinsic dipole, as discussed in our companion letter~\cite{Ferreira:2020aqa}. The up-coming ground-based experiments will reach smaller scales than Planck, and in particular CMB-S4 will probe the CMB at $\ell > 3000$, improving the precision in both estimators. But the Doppler couplings will remain harder to measure than the aberration couplings. In fact, the improvements in Doppler by going all the way to $\ell_{\rm max} = 4000$ will be only roughly a factor of 2, whereas aberration, which is already more precisely measured, should improve by a factor of~3.  Finally, since our estimators are based on the first-order Taylor expansion of the Doppler and aberration kernels, all information is assumed to be contained in the $\ell, \ell+1$ correlations. For $\ell < 3000$ this assumption was shown to be a good approximation in~\cite{Notari:2011sb}. For smaller scales this will need to be investigated in more detail.

\bigskip

\section*{Acknowledgements}

We would like to thank Hans Kristian Eriksen,  Alessio Notari, Douglas Scott, Soumen Basak,  Maude Le Jeune and Suvodip Mukherjee for useful discussions. We also thank the anonymous referee for helpful suggestions, in particular on an alternative CCP pipeline. PSF is supported by the Brazilian research agency CAPES (Coordenação de Aperfeiçoamento de Pessoal de Nível Superior). MQ is supported by the Brazilian research agencies FAPERJ and CNPq (Conselho Nacional de Desenvolvimento Científico e Tecnológico).

\bibliography{cmb}

\appendix

\section{Expected uncertainties in each multipole for aberration and Doppler}
\label{app-error-by-ell}

\begin{figure*}
    \includegraphics[trim={0.5cm 0cm 0cm 0cm}, clip, width=2\columnwidth]{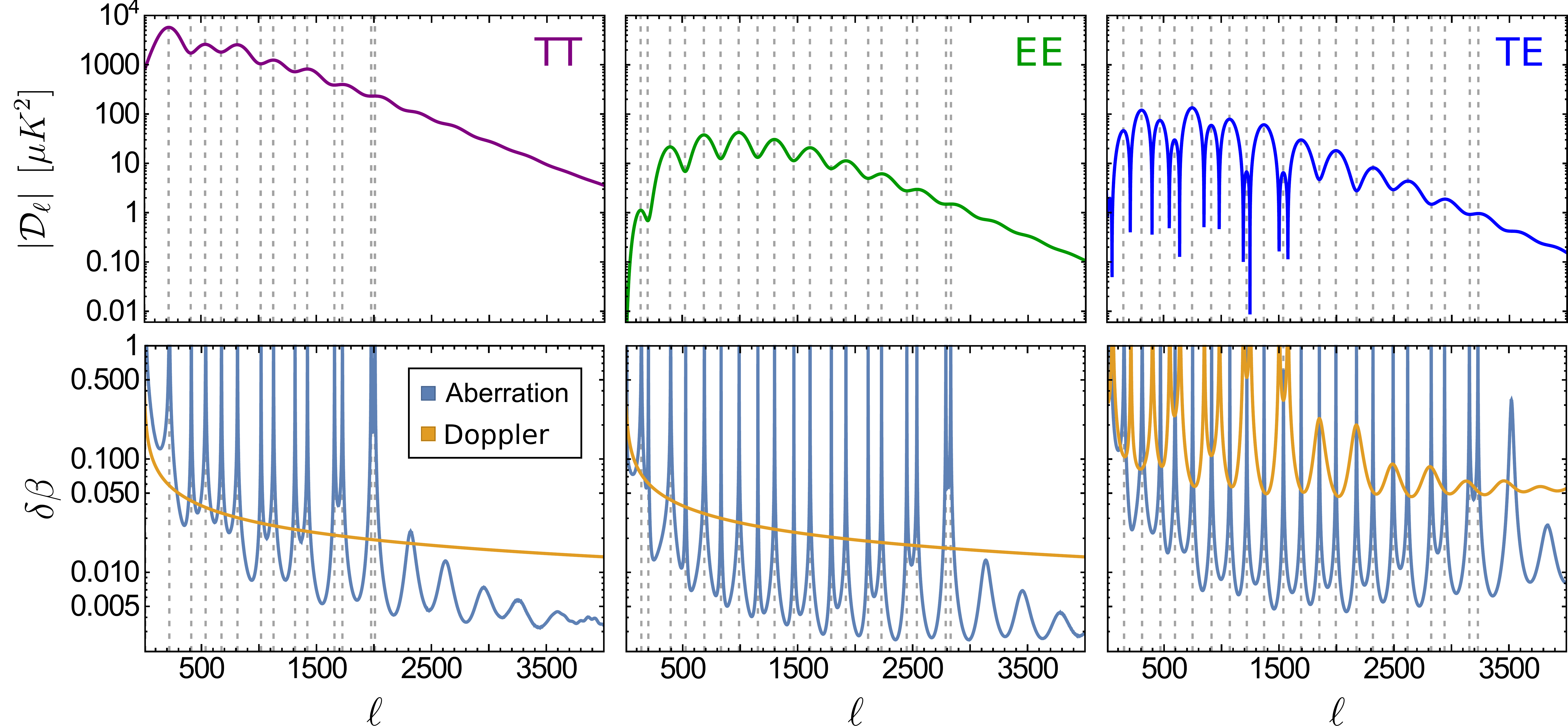}
    \caption{Angular power spectra (Top) and estimated uncertainties (Bottom) for aberration and Doppler in the ideal case for the $TT$, $EE$, and $TE$ two-point correlations. Here we show the dimensionless values of $\delta \beta$ instead of $c\delta \beta$. Vertical dashed lines indicate the extrema of $\mathcal{D}_{\ell}$, in which the aberration effect produces no signal. The Doppler uncertainties are approximately independent of the shape of the spectrum for $TT$ and $EE$, and are proportional to $(2\ell+1)^{-1/2}$.
    \label{fig:errors-by-ell}}
\end{figure*}

In order to better understand the uncertainties in both Doppler and aberration estimators in each multipole we follow a similar reasoning to the one included in~\cite{Burigana:2017bxl}. The idea is to compute the signal-to-noise ratio in each $\ell$ by first noting that, as shown in~\cite{Amendola:2010ty,Notari:2011sb}, the fractional uncertainty in the estimator of $ \left<a_{\ell m}^{X}~a_{(\ell+1)m}^{Y\ast}\right>$ is given by
\begin{equation}\label{eq:delta-beta}
    \left. \frac{\delta \beta}{\beta}\right|_{XY} \simeq \left[\sum_{\ell}\sum_{m=-\ell}^\ell \frac{\left< a_{\ell m}^{X}~a_{(\ell+1)m}^{Y\ast}\right>^{2}} {{\mathfrak C}^{XX}_{\ell} {\mathfrak C}^{YY}_{\ell+1}} \right]^{-1/2}\,.
\end{equation}
The expected value above involve a sum over all $\ell$s and $m$s. Since $m$ enters only through the coefficients in Eq.~\eqref{eq:Glm} we can make simplify things considerably by first performing an average on $m$ on the coefficients:
\begin{equation}\label{eq:glm-average}
    \sum_m \big[G_{\ell m}\big]^2 \simeq 0.408^2 (2\ell + 1).
\end{equation}
If we replace $C_{\ell+1} \rightarrow C_{\ell} + \dd C_{\ell}/\dd \ell$, and use Eqs.~\eqref{eq:almab} and \eqref{eq:almdopp} we get the following approximation for the aberration and Doppler signal squared in each multipole $\ell$:
\begin{align}\label{eq:average-sum-m}
    \sum_m & \left<a_{\ell m}^{X}~a_{(\ell+1)m}^{Y\ast}\right>^2
    \,=\,  0.408^2 (2\ell +1) \beta^2 \nonumber \\
    &\qquad \left[ (2a-2d) C_{\ell}^{XY} + (a\ell+d) \frac{\dd C_{\ell}^{XY}}{\dd \ell}
    \right]^2\,,
\end{align}
where $a$ and $d$ are boolean variables such that $a=1$ ($d=1$) when including aberration (Doppler), and zero otherwise. This can be simplified using the approximation that $\dd C_{\ell}/\dd \ell \ll C_{\ell}$ and by noting that
\begin{equation}
    \frac{\dd \ln \mathcal{D}_{\ell}^{XY}}{\dd \ln \ell} = \frac{\dd \ln C_{\ell}^{XY}}{\dd \ln \ell} + \frac{2\ell+1}{\ell +1} \simeq \frac{\dd \ln C_{\ell}^{XY}}{\dd \ln \ell} + 2\,.\smallskip
\end{equation}
In the last passage we just assumed $\ell \gg 1$. We thus arrive at
\begin{align}\label{eq:average-sum-m-simp}
    \sum_m & \left<a_{\ell m}^{X}~a_{(\ell+1)m}^{Y\ast}\right>^2
    \,=\,  0.408^2 (2\ell +1) \beta^2 \nonumber \\
    &\qquad \big(C_{\ell}^{XY}\big)^2 \left[ -2d + a \frac{\dd \ln \mathcal{D}_{\ell}^{XY}}{\dd \ln \ell}
    \right]^2\,.
\end{align}

This result means that the aberration signal is proportional to the logarithm derivative of $\mathcal{D}_{\ell}$, which means for instance that at the extrema of $\mathcal{D}_{\ell}$ there is no aberration signal. In the particular case in which $X=Y$ (i.e., for $TT$ and $EE$, but not for $TE$) and in the ideal case where there is no noise or masks so that ${\mathfrak C}^{XX}_{\ell} = C^{XX}_{\ell}$ the $C_{\ell}^{XX}$ term in front of the brackets cancel and we arrive at the following uncertainties in $\beta^{\rm{A,D}}$ in any given multipole $\ell$:
\begin{align}\label{eq:betaAD-error-approx}
     \delta \beta^{\rm A}(\ell)\Big|_{XX} &\simeq 2.451
     \big[2\ell + 1\big]^{-1/2}
     \left( \frac{\dd \ln \mathcal{D}_\ell^{XX}}{\dd \ln \ell} \right)^{-1}, \\
     \delta \beta^{\rm D}(\ell)\Big|_{XX} &\simeq 1.225
     \big[2\ell + 1 \big]^{-1/2} .
\end{align}
Interestingly, this means that for $X=Y$ the Doppler signal is \emph{independent} of the shape of the angular power spectrum. Figure~\ref{fig:errors-by-ell} illustrates that this simple behaviour of both uncertainties is indeed found in the full estimator (without averaging over $m$) in the ideal, noiseless case.

Another curious conclusion from Eq.~\eqref{eq:average-sum-m-simp} is that for a traditional Boost estimator, in which $a=d=1$, there is a partial cancellation of the total signal whenever $\dd \ln \mathcal{D}_{\ell}^{XY} / \dd \ln \ell$ is positive. This means that in some multipole ranges the boost estimator is less precise than the aberration estimator alone, even though aberration is just part of the effect. In practice, this partial cancellation is more important in  the $EE$ case, as depicted in Figure~\ref{fig:planck-forecasts}, where for $\ell_{\rm max} < 1000$ the Boost estimator exhibits less precision than the aberration one.

\label{lastpage}

\end{document}